\documentclass{ws-ijmpa}
\usepackage[super,compress]{cite}
\usepackage{graphicx}
\usepackage{hyperref}
\usepackage{url}
\hypersetup{colorlinks,urlcolor=black,citecolor=black,linkcolor=black,filecolor=black}
\usepackage{breakurl}
\usepackage{graphicx,amsmath,amssymb,bm}
\usepackage{color}
\usepackage[utf8]{inputenc}
\usepackage{hyperref}
\usepackage{amssymb}
\usepackage{graphics}
\usepackage{multirow}
\usepackage{epsfig}
\usepackage{rotating}
\usepackage{longtable} 
\usepackage{url}
\usepackage{pdflscape}
\usepackage{cancel}
\pdfoutput=1
\usepackage{csquotes}
\usepackage{placeins}
\usepackage{scalerel}
\usepackage{widetext}
\usepackage{color}

\begin{document}

\markboth{Richard H.~Benavides, Luis Mu\~noz, William A.~Ponce, Oscar Rodr\'iguez, Eduardo Rojas }{Paper}

%
\catchline{}{}{}{}{}
%

\title{Electroweak couplings and LHC constraints\\ on alternative $Z^{\prime}$ models  in $E_6$ }
\author{Richard H.~Benavides$^{a}$, Luis Mu\~noz$^{a}$, William A.~Ponce$^{b}$, Oscar Rodr\'iguez$^{a,b}$, Eduardo Rojas$^{c,b}$}

\address{$^{a}$ Facultad de Ciencias Exactas y Aplicadas, Instituto Tecnol\'ogico Metropolitano,
\\Calle 73 No 76 A - 354 , V\'ia el Volador, Medell\'in, Colombia
}
\address{$^{b}$ Instituto  de F\'isica, Universidad de Antioquia,\\Calle 70 No. 52-21, Medell\'in, Colombia
}
\address{$^{c}$
Departamento de F\'\i sica, Universidad de Nari\~no,\\ 
A.A. 1175, San Juan de Pasto, Colombia}
\maketitle

\begin{history}
\received{Day Month Year}
\revised{Day Month Year}
\end{history}

\newcommand{\ie}{{\emph i.e.,\ }}
\newcommand{\eg}{\emph{e.g.,\ }}

 \newcommand{\cosx}{\cos\theta}
 \newcommand{\sinx}{\sin\theta}
 \newcommand{\tanx}{\text{T}_\theta}
 \newcommand{\cosxs}{\cos^2\theta}
 \newcommand{\sinxs}{\sin^2\theta}
 \newcommand{\cosy}{\cos\phi}
 \newcommand{\siny}{\sin\phi}
 \newcommand{\tany}{\tan\phi}
 \newcommand{\cosz}{\cos\omega}
 \newcommand{\sinz}{\sin\omega}
 \newcommand{\tanz}{\tan\omega}

\newcommand{\g}{g_L\text{T}_W}
\newcommand{\y}{g_b}
\newcommand{\z}{g_c}
\newcommand{\s}{\text{S}_\omega}
\newcommand{\w}{\text{C}_\omega}
\newcommand{\p}{\dfrac{g^2_L\text{T}^2_W}{\alpha}}
\newcommand{\q}{\alpha}
\newcommand{\U}{\dfrac{1}{\beta}}
\newcommand{\V}{\beta}

\newcommand{\Q}{Z}

\newcommand{\ha}{\hat{\alpha}_c}
\newcommand{\hb}{\hat{\beta}}
\newcommand{\haa}{\hat{\alpha}_a}
\newcommand{\hab}{\hat{\alpha}_b}

\newcommand{\gp}{g_{Y}}
\newcommand{\gps}{g^{2}_{Y}}
\newcommand{\gpp}{g^{\prime\prime}}
\newcommand{\gzp}{g_{Z^\prime}}
\newcommand{\gzpp}{g_{Z^{\prime\prime}}}
\newcommand{\zp}{Z^{\prime}}
\newcommand{\zpp}{Z^{\prime\prime}}
\newcommand{\pp}{\prime\prime}
\newcommand{\ovl}{\overline}

\begin{abstract}
We report the most general expression for the chiral charges of a  $Z'$ gauge  boson coming from an  $E_6$ unification model, 
as a function  of the electroweak parameters and  the charges of the $U(1)$ factors in the chain of subgroups.  
These charges are valid for an arbitrary Higgs sector and only depend on the branching rules  
of the $E_6$ fundamental representation and the corresponding rules for the fermionic representations of their subgroups.
By assuming $E_6$ unification, the renormalization group equations~(RGE) allow us to calculate the electroweak 
parameters at low energies for most of the chains of subgroups in $E_6$.
From RGE  and unitary conditions,  we show  that at low energies there must be 
a mixing between the gauge boson  of the  standard model hypercharge and the $Z'$.
From this, it is possible to  delimit the preferred 
region in the parameter space for a  breaking pattern in $E_6$. 
In general,  without unification, it is not viable to determine this region; however, for some models
and under certain assumptions, it is possible to limit the corresponding parameter space. 
By using the most recent 
upper limits  on the cross-section  of extra gauge vector bosons  $Z'$ decaying into dileptons
from the  
ATLAS data  at 13~TeV 
with accumulated luminosities of  36.1~fb$^{-1}$ and  13.3~fb$^{-1}$,
we  report the   95$\%$ C.L. lower limits  on  the  $Z'$ mass for the typical $E_6$ benchmark models.   
We also show the contours in the 95\% C.L. of the  $Z'$ mass bounds for the entire  parameter space of $E_6$.


\end{abstract}

\ccode{PACS numbers:}


\section{Introduction  \label{intro}}
From a group theory point of view, there are several ways to break the $E_6$ symmetry down to the standard model~(SM) one.  
Although some of the breaking patterns have been explored  in the literature so far, a systematic study 
of the phenomenology for all the alternative ways has not been done as far as we know.
 In general, intricate models are not appealing. A way to look for new models with a moderate fermion content   
is to consider alternative versions of the  models already  known in 
the literature~\cite{Barr:1981qv,Robinett:1982tq,Witten:1985xc,Ma:1986we,Ma:1995xk,Martinez:2001mu,Rojas:2015tqa,Mantilla:2016sew,Huang:2017uli,
CarcamoHernandez:2017owh}.
Our work represents a first step in this  direction.
One of the first alternative models was ``flipped $SU(5)$''~\cite{Barr:1981qv,Derendinger:1983aj},
which produces a symmetry breaking for $SO(10)$  down to $SU(5)\otimes U(1)$, 
where the $U(1)$ factor contributes to the electric charge, 
and as such, its basic predictions for $\sin^2 \theta_W$  and the proton decay 
are known to be different from those of $SU(5)$~\cite{Barr:1981qv}. 
An alternative model   for the  flipped $SU(5)$  is   $SO(10) \otimes U(1)_{N}$,  
where the right-handed neutrino has zero charge under the  $U(1)_N$ 
group  allowing   a Majorana mass term~\cite{Ma:1995xk,King:2005jy}. 
The alternative versions of the left-right model are also well known 
in the literature~\cite{Ma:1986we,Rodriguez:2016cgr},  as in the latter case, one of these 
models  allows for  a right-handed neutrino  component   effectively inert~\cite{Ma:1986we}.

The alternative models have been useful in the study of the grand unified theories~(GUTs) phenomenology,
 for example, for  the  $E_6$ subgroup $SU(2)_R\otimes SU(6)$ and some of 
its three alternative versions, the gauge mediated proton decay operators are suppressed at leading order due to the special 
placement of matter  fields in
the  unified multiplets~\cite{Dimopoulos:1985xs,Rizos:1997pe,Shafi:1998jf,Faraggi:2014bla,Huang:2017uli,Dong:2017zxo}.
 
Heavy neutral gauge bosons are a generic prediction of many types of new physics beyond the SM. 
In addition, these  extra $U(1)'$ symmetries serve as an important model-building tool~\cite{Benavides:2016utf} 
(for example, to suppress strongly constrained processes)
giving rise,  after spontaneous $U(1)'$ symmetry breaking,  to physical $Z'$ vector bosons. 
Thus, with the upgrade of the luminosity  and the energy of the LHC, 
there exists a real possibility for the on-shell production of a $Z'$ boson~\cite{Langacker:2008yv,Salazar:2015gxa}. 

All representations of the $E_6$ gauge group~\cite{Gursey:1975ki,Achiman:1978vg} are anomaly-free and 
the fundamental {\bf 27}-dimensional representation is chiral and can accommodate a full SM fermion generation.
As a consequence, $E_6$-motivated $Z'$ bosons arise naturally 
in many popular extensions of the SM~\cite{Langacker:2008yv,Robinett:1982tq,London:1986dk,Camargo-Molina:2017kxd},
both in  top-down and  bottom-up constructions.
Some of the $E_6$ subgroups, such as the original unification groups $SU(5)$ and $SO(10)$,
and the gauge group of the left-right symmetric models $SU(4)_C\otimes SU(2)_L\otimes SU(2)_R$, play 
central roles in some of the best motivated extensions of the SM.   
Furthermore, the complete $E_6$-motivated $Z'$ family of models appears in a supersymmetric bottom-up approach 
exploiting a set of widely accepted theoretical and phenomenological requirements~\cite{Erler:2000wu}.
The one-parameter $Z'$ families in reference~\cite{Carena:2004xs}, denoted as  ${\bf 10} + x {\bf \bar{5}}$, $d - x u$ and $q + x u$,
where ${\bf 10}$ and ${\bf \bar{5}}$ are $SU(5)$ representations,  can also be discussed within the $E_6$ framework~\cite{Erler:2011ud}.

For all these reasons there is an expectation that an $E_6$ Yang-Mills theory,
or a subgroup of $E_6$ containing the SM in a non-trivial way,
might be part of a realistic theory~\cite{Slansky:1981yr}.
If a heavy vector boson is seen at the LHC or at an even more energetic collider in the future,
aspects of the $E_6$ symmetry group will be central to the discussion of what this resonance might be telling us 
about the fundamental principles of nature.
  
The discrimination between $Z'$ models could be challenging at the LHC due to the small number 
of high resolution channels at hadron colliders.
Another reason why the determination of the underlying symmetry structure is not straightforward 
is that the mass eigenstate of the $Z'$ is, in general, a linear combination of some of the underlying $Z'$ charges
with the ordinary $Z$ boson of the SM.
Hence, it is useful to reduce the theoretical possibilities or at least to have a manageable setup.
This work represents an attempt in this direction and serves to spotlight a few tens of models in the two-dimensional   
space of $E_6$-motivated $Z'$ models.

All the $E_6$ breaking patterns and branching rules have been tabulated in Ref.~\cite{Slansky:1981yr}. 
In references~\cite{Robinett:1982tq,Rojas:2015tqa} all the chains of subgroups were tabulated.
The aim of the present work is to set the impact of the latest 
LHC constraints on the possible embeddings of the SM 
in the subgroups of  $E_6$. 

It is important to remark that many interesting phenomenological models appear 
in a natural way in $E_6$ breaking patterns, such as the 
 proton-phobic, $Z_{\not{p}}$,  neutron-phobic, $Z_{\not{n}}$,~(vector  bosons which at zero momentum transfer do not  couple 
to protons and neutrons, respectively),  leptophobic $Z_{\not{L}}$~(with zero couplings to leptons) and  
vector bosons from supersymmetric models, as for example the $ Z_N$ model~\cite{Ma:1995xk,King:2005jy}, {\em etc}.
We will show a more complete list later.

The study reported here is  a continuation of the analysis started in Ref.~\cite{Rojas:2015tqa}
 where   the quantum numbers of the abelian gauge groups 
in alternative chains of subgroups of $E_6$ were calculated. Several of the subgroups 
shown there are well known in the literature; however, as far as we know, 
the phenomenology of many of these models have not been studied.  
Of particular importance for the electroweak constraints are the $Z'$ chiral
charges of the SM fermions which depend on the  chosen  chain of subgroups. In the present work, 
we show the general expression for these charges and determine the preferred region 
in the parameter space for some breaking patterns. We also establish that the mixing 
between the $Z'$ charges and the SM hypercharge is a measure of the deviation of the parameter 
space at low energies respect to their unification values.  We demonstrate that
the presence of this mixing stems from the gauge coupling splitting at low energies. 

 The paper is organized as follows:
 In Section~\ref{sec:general} we derive general expressions 
 for the electroweak~(EW) charges of a $Z^{\prime}$ in $E_6$  
 as a function of the mixing angles and the charges of an arbitrary  $U(1)$ in $E_6$.
 In section~\ref{sec:kmixing} we show that even for a group with orthonormal charges at low energies   there is a kinetic mixing due to the splitting of the gauge coupling constants.
 In section~\ref{sec:models} we revise  the existing literature 
 about models based on  $E_6$ subgroups and their embeddings.  
 By assuming $E_6$ unification the renormalization group equations~(RGE) 
 allow us to determine the parameter space of the $Z'$ associated with some of these models.   
For this purpose, we take the expressions for the mass scales and couplings of the Robinett and Rosner~(RR) work~\cite{Robinett:1982tq}.
In this section, we also point out the existence of non-trivial models which, 
to the best of our knowledge, have not been studied in the literature. 
In section~\ref{sec:lem} we delimit  the parameter space when we put aside the unification hypothesis
as it  usually happens for effective models at low energies.
In section~\ref{sec:constraints} the 95\% C.L. exclusion limits   on the neutral boson masses   
for the entire $E_6$-motivated $Z'$ parameter space are shown.

\section{General expressions}
\label{sec:general}
 Owing to the fact that the rank of  $E_6$  is 6, for  chains of subgroups with regular
embeddings~(those preserving the rank) the most general form of the group associated with 
the low-energy effective model is~\cite{Slansky:1981yr,Robinett:1982tq} $SU(3)_C\otimes SU(2)_L\otimes\prod_{\kappa} U(1)_{\kappa}$, 
with $\kappa=a,b,c$; where the $U(1)$ factors come from the chains of subgroups of $E_6$.  
In order to reproduce the  $SU(3)_C\otimes SU(2)_L\otimes U(1)_Y$ symmetry of the SM
it is necessary that the SM hypercharge $Y$ be a linear combination of the $U(1)_{k}$ charges 
 $Q_\kappa$.    If $g$ is the $SU(2)_L$ coupling 
constant and $A^\mu_{3L}$ the gauge boson associated with the third component of the weak isospin,
then the neutral current Lagrangian $\mathcal{L}_{NC}$ for the most general case is
\begin{equation}\label{eq:lagrangian}
-\mathcal{L}_{NC} = gJ^\mu_{3L} A_{3L\mu} + g_aJ^\mu_a A_{a\mu} + g_bJ^\mu_b A_{b\mu} + g_cJ^\mu_c A_{c\mu} \ ,  
\end{equation}
where   $g_\kappa$ and $A^\mu_\kappa$ represent the 
gauge coupling constant and the gauge field associated with the $U_\kappa (1)$ symmetry, respectively.
The fermion currents $J^\mu_\kappa$ are given by
\begin{equation}\label{eq:Eq2}
J^\mu_\kappa = \sum_{f}\overline{f}\gamma^\mu [\epsilon_L^{\kappa}(f) P_L + \epsilon_R^{\kappa}(f) P_R] f,
\end{equation}
where $f$ runs over all fermions in the ${\bf 27}$ representation of $E_6$,  which is  the fundamental representation. 
 The chirality projectors are defined as usual \ie  $P_{L,R} = (1\pm \gamma^5)/2$ and $f_{L,R}=P_{L,R}f$.
The chiral  charges  are  $\epsilon_L (f)= Q_k (f_L) $ and   $\epsilon_R (f)=- Q_k (f^c_L) $.
The  $U(1)_\kappa$ charges satisfy the  $E_6$  normalization condition $\sum_{f\in \mathbf{27}}Q^2_\kappa(f)=3$~(see Table \ref{tab:chiral} in appendix~\ref{sec:tables}).  
As a consequence of this, the electric charge operator $Q_{em}$ is given by
\begin{equation*}
Q_{em} = T_3 + Y=T_3 + \sqrt{\frac{5}{3}}Q_Y^{E_6}\ ,
\end{equation*}
where $T_3$ is the third component of weak isospin and $Q_Y^{E_6}$ is the $E_6$ normalized SM hypercharge. 
\\\\
By means of an orthogonal transformation $\mathcal{O}$ we can pass from the  gauge interaction  basis to the basis in which one
of the fields can be identified with the SM hypercharge $B^\mu$ associated with the $U(1)_Y$ symmetry. 
If we define such a rotation through~\footnote{
The absence of  mixing  between $A_{3L}^\mu$ and the other fields 
is related to the strong constraints on the  $Z$ and $Z'$ mixing angle by 
low energy experiments\cite{Erler:2009jh}, consequently, the only   mixing between the  hypercharge and 
 $A_{3L}^\mu$ is  parametrized by the Weinberg  angle $\theta_W$.
} 

\begin{equation}\label{rotation1}
\begin{pmatrix}
A^\mu_{3L} \\
A^\mu_a \\
A^\mu_b \\
A^\mu_c 
\end{pmatrix} =
\mathcal{O} 
\begin{pmatrix}
A^\mu_{3L} \\
B^\mu \\
Z^{\prime\mu} \\
Z^{\pp\mu} 
\end{pmatrix}
=
\begin{pmatrix}
1 &0 &0 &0 \\
0 &O_{11}  &O_{12}  &O_{13}  \\
0 &O_{21}  &O_{22}  &O_{23}  \\
0 &O_{31}  &O_{32}  &O_{33}  
\end{pmatrix}
\begin{pmatrix}
A^\mu_{3L} \\
B^\mu \\
Z^{\prime\mu} \\
Z^{\pp\mu} 
\end{pmatrix}\ ,
\end{equation} 
then the  Lagrangian~(\ref{eq:lagrangian}) can be written as 
\begin{equation}
-\mathcal{L}_{NC} = gJ^\mu_{3L} A_{3L\mu} + \gp J^\mu_Y B_{\mu} + g_{\zp}J_{\zp}^\mu Z^{\prime}_{\mu} + g_{\zpp}J_{\zpp}^\mu Z^{\pp}_{\mu} \ .  
\end{equation}
In order to keep invariant the Lagrangian, the currents must transform with the same orthogonal matrix 
\begin{align}
\gp J^\mu_Y                &= g_a J^\mu_aO_{11} + g_b J^\mu_b O_{21} + g_c J^\mu_c O_{31}\ , \label{Eq1}\\ 
g_{\zp} J_{\zp}^\mu    &= g_a J^\mu_aO_{12} + g_b J^\mu_b O_{22} + g_c J^\mu_c O_{32}\ , \label{Eq1b}\\
g_{\zpp}J_{\zpp}^\mu   &= g_a J^\mu_aO_{13} + g_b J^\mu_b O_{23} + g_c J^\mu_c O_{33}\ . \label{Eq1c} 
\end{align}
The exact expression for the orthogonal matrix is given in  appendix~\ref{sec:rotation}.
In order to obtain the SM as an effective theory at low energies, the breaking $U_a(1)\otimes U_b(1)\otimes U_c(1) \longrightarrow U(1)_Y$ must take place. If so, 
it is possible to find three real coefficients $k_a$, $k_b$ and $k_c$ such that
\begin{align}\label{EqY}
 Y= \sqrt{\frac{5}{3}}Q_Y^{E_6} =& k_a Q_a + k_b Q_b + k_c Q_c\ . 
\end{align}
From Eqs.~(\ref{eq:Eq2}) and (\ref{EqY}) we obtain for the currents  the relation 
\begin{align}\label{EqY2}
J_Y^\mu=  k_a J_a^\mu + k_b J_b^\mu + k_c J_c^\mu\ .
\end{align}
In Tables \ref{table:a} to \ref{table:d}~(in appendix~\ref{sec:tables}) we have reported the values of $k_{\kappa}$ for the  models considered in this work. 
By comparing  Eq.~(\ref{EqY2}) with Eq.~(\ref{Eq1}), we get  the following expressions:
\begin{align}\label{Eq4}
g_aO_{11} = k_a\gp\ , \hspace{0.5cm} g_bO_{21} = k_b\gp\ , \hspace{0.5cm} g_cO_{31} = k_c\gp\ ,
\end{align}
which, along with the orthogonal condition $O_{11}^2 + O_{21}^2 + O_{31}^2 = 1$, impose a constraint on the $g_a$, $g_b$ and $g_c$ coupling constants, namely:
\begin{equation}\label{Eq5b}
\left(\frac{k_a}{g_a}\right)^2 + \left(\frac{k_b}{g_b}\right)^2 + \left(\frac{k_c}{g_c}\right)^2 = \left(\frac{1}{\gp}\right)^2\ .
\end{equation}
From these expressions and  the explicit form of the rotation matrix $\mathcal{O}$ (see \ref{sec:rotation}) we get
the $Z'$ chiral charges~(see \ref{sec:zprimecharges})
\begin{align}\label{eq:zprimecharges}
g_{Z^\prime}\epsilon^{Z^\prime}_{L,R}=&\ \ \ A_{L,R}\cosx+ B_{L,R}\sinx\ , \\
\gzpp       \epsilon^{\zpp}_{L,R}    =&-A_{L,R}\sinx+ B_{L,R}\cosx\ , 
 \end{align}
  where 
\begin{align}
 A_{L,R}=& 
  \hspace{0.3cm}
 \frac{g_c}{\ha}\left(\dfrac{k_a\epsilon^b_{L,R}}{\hb}-\hb k_b \epsilon^a_{L,R} \right),  \\
  B_{L,R}=&
    \gp\left(-\dfrac{k_a \epsilon^a_{L,R}+k_b \epsilon^b_{L,R}}{\ha}k_c+\ha \epsilon^c_{L,R}\right)\ ,
\end{align}
and $\theta$ is an angle  of the rotation matrix $\mathcal{O}$ which can take any value between 
$-\pi$ and $\pi$.  Here  
$\hat{\alpha}_{c} 
=g_c\sqrt{\frac{k_a^2}{g_a^2}+\frac{k_b^2}{g_b^2}}
=\sqrt{\frac{g_c^2}{g^2}\cot^2 \theta_W-k_c^2}$,
and  $\hb = \frac{g_a}{g_b}$.
In order to have the chiral charges properly normalized in $E_6$ we define 
\begin{align}\label{eq:gzp}
\gzp\equiv & \frac{1}{{\ha\hb }} 
\bigg(
\cosxs\z^2(k_a^2+\hb^4k_b^2)-2\hb(1-\hb^2)\z\gp k_a k_b k_c\cosx\sinx\\
&+\hb^2\gps \left((k_a^2+k_b^2)k_c^2+\ha^4\right)\sinxs
\bigg)^{1/2}
\ , 
\end{align}
which reduces to the Georgi-Glashow well known result $\sqrt{\frac{5}{3}} \gp=\sqrt{\frac{5}{3}} g\tan\theta_W$ for $g_c=g_a=g_b$.
 These charges reproduce the electroweak charges of  trinification and the 
left-right symmetric model which are well known in the literature~(for additional references look into our previous work~\cite{Rodriguez:2016cgr}).
Since  $g_{Z^\prime}\epsilon^{Z^\prime}_{L,R}(\theta+\pi/2)= g_{Z''}\epsilon^{Z''}_{L,R}(\theta)$,
the parameter space associated with the $Z''$ boson is the same as that of the $Z'$ boson.

\section{Kinetic mixing from  gauge coupling splitting}
\label{sec:kmixing}
Because all the  generators $Q_{ij}^a$ associated with the neutral currents  can be diagonalized simultaneously the corresponding fields can be written as 
$A^{\mu}=A^{\mu a}T^{a}_{ij}= A^{\mu a}Q^{a}(i)\delta_{ij}$,  
where   $Q^{a}(i)$ stands for the charge of the $i$-fermion  in the fundamental representation. 
For these fields the most general lagrangian is given by
\begin{align}
\text{Tr}[F_{\mu\nu}F^{\mu\nu}]=&\text{Tr}[F_{\mu\nu}^aT^{a}F^{\mu\nu b}T^{b}]\notag\\
                        F_{\mu\nu}^{a}F^{\mu\nu b}\sum_{i,j}Q^{a}(i)\delta_{ij}Q^{b}(j)\delta_{ji}= &
                        F_{\mu\nu}^{a}F^{\mu\nu b}\sum_{i}Q^{a}(i)Q^{b}(i).
\end{align}
When  $i$ runs over the  fermions in a  multiplet of a simple group (or a semisimple group that comes from the breaking of  a simple Lie group) the 
charges are orthonormal 
\begin{align}
\sum_{i}Q^{a}(i)Q^{b}(i)=  
\sum_{i}\epsilon^{a}(i)\epsilon^{b}(i)= \delta_{ab}\ .
\end{align}
It is not possible 
to generate a  kinetic mixing term to tree level 
because $F^{\mu\nu a}$   transforms  with an orthogonal matrix; 
however, at low energies it is possible to generate a kinetic mixing by 
one-loop corrections~\cite{Holdom:1985ag,Babu:1996vt,Babu:1997st,Langacker:2008yv}.
As we will show, a  source of kinetic mixing at low energies 	
is the splitting of the values of the coupling strengths.
By unitarity the currents should 
transform in the same way as the fields; if we transform from the group basis
to a basis where one of the fields corresponds to the  vector 
field   associated with the SM hypercharge $B_\mu$, the corresponding expression for the currents is
\begin{equation}\label{rotation2}
\begin{pmatrix}
    g J^\mu_{3L} \\
  \gp J^\mu_{Y} \\
 \gzp J^{\prime\mu} \\
\gzpp J^{\pp\mu} 
\end{pmatrix}\ 
=
\mathcal{O}^T 
\begin{pmatrix}
  g J^\mu_{3L} \\
g_a J^\mu_a \\
g_b J^\mu_b \\
g_c J^\mu_c 
\end{pmatrix} 
=
\begin{pmatrix}
1 &0 &0 &0 \\
0 &O_{11}  &O_{12}  &O_{13}  \\
0 &O_{21}  &O_{22}  &O_{23}  \\
0 &O_{31}  &O_{32}  &O_{33}  
\end{pmatrix}
^{T}
\begin{pmatrix}
g   J^\mu_{3L} \\
g_a J^\mu_a \\
g_b J^\mu_b \\
g_c J^\mu_c 
\end{pmatrix}\ . 
\end{equation} 
From the definitions
$ \gp J_Y^{\mu}(f) =
  \sum_{\kappa} \mathcal{O}_{1\kappa}g_{\kappa}J_{\kappa}^{\mu}(f)$ ,
$\gzp J_{\zp}^{\mu}=\sum_{\lambda}\mathcal{O}_{2\lambda}g_{\lambda}J_{\lambda}(f)^{\mu}$,
and the Eq.~\ref{eq:Eq2} we obtain the expressions 
\begin{align}
 \gp  Y=&\sum_{\kappa} \mathcal{O}_{1\kappa}g_{\kappa}\epsilon^{\kappa}(f)\ ,\\
 \gzp \epsilon_{\zp}=&\sum_{\lambda}\mathcal{O}_{2\lambda}g_{\lambda}\epsilon^{\lambda}(f)\ .
\end{align}
  By taking  the dot product
of the SM hypercharge $gY$ and 
the $Z'$ charges $g_{\zp}{\bm\epsilon}_{\zp}$   we obtain 
\begin{align}\label{eq:orth}
\gp {\bf Y}\cdot \gzp {\bm \epsilon}_{\zp} \equiv & \sum_{f\in 27} \gp Y(f) \gzp \epsilon_{\zp}(f)\notag\\
                              =& \sum_{\kappa,\lambda} \sum_{f\in 27}\mathcal{O}_{1\kappa}g_{\kappa}
                               \epsilon^{\kappa}(f)\mathcal{O}_{2\lambda}g_{\lambda}\epsilon^{\lambda}(f)\\
                              =& 3\sum_{\kappa} \mathcal{O}_{1\kappa}\mathcal{O}_{\kappa2}^{T}g_{\kappa}^2\ , 
\end{align}
where $\mathcal{O}$ is the rotation matrix~(\ref{rotation1}).
Here we made use of the $E_6$ orthonormality relation $\sum_{f\in 27}\epsilon^{\kappa}(f)\epsilon^{\lambda}(f)=3\delta_{\kappa\lambda}$
between  the  $U(1)_{\kappa}$ charges that  come from  a chain of subgroups.
By assuming that the three couplings are identical  $g_a=g_b=g_c$ we obtain 
$\gp {\bf Y}\cdot \gzp {\bm \epsilon}_{\zp}= 3g_{a}^2\sum_{\kappa} \mathcal{O}_{1\kappa}\mathcal{O}_{\kappa2}^{T}=  
                 3g_{a}^2 \delta_{12}= 0$, otherwise
\begin{align}
 {\bf Y}\cdot {\bm \epsilon}_{\zp}=\sum_{f\in 27}  Y(f)  \epsilon_{\zp}(f)\neq 0\ .
\end{align}

\begin{figure}[h]
\centering 
\begin{tabular}{c}
\includegraphics[scale=0.7]{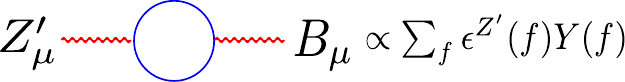}  
\end{tabular}
\caption{$Z_{\mu}'$-$B_{\mu}$ kinetic mixing.}
\label{fig:km}
\end{figure}
This result shows that the orthonormality of the 
SM hypercharge and the $Z'$ charges  is only guaranteed  when all the three couplings 
are equal  as it happens in unification; for the remaining cases  a kinetic mixing is generated by the one-loop diagram in figure~\ref{fig:km}
(even for complete fermion representations).

In general, at low energies  the gauge couplings $g_{\kappa}$ are different each other due to the  RGE; 
thus, as will be shown below, the $Z'$ charges   associated with a chain of subgroups in $E_6$ are no longer orthonormal to the SM hypercharge. 
 It is important to notice that the 2-loop corrections are important for the RGE since they modify
in a considerable way the mass unification scales; however, for several models, 
unification does not impose relations between the SM electroweak couplings 
in such  a way that the consistency of the model does not depend on  high order corrections  and the SM
values for the $\alpha_i=g_i^2/(4\pi)$  can be considered as input parameters. 
 Under these conditions,  the 1-loop  coupling strengths~\cite{Robinett:1982tq} $g_\kappa$
 associated to the extra $U(1)$  abelian symmetries differs in just a few percent respect to the 2-loop result at the electroweak scale, 
since the bondary condition on the SM parameters is imposed at the same energy scale~\cite{delAguila:1988jz}.

The phenomenological consequences of the $B_\mu-Z_\mu^{\prime}$ kinetic mixing is a modification of the $Z$ charges~\cite{Babu:1996vt,Babu:1997st,Langacker:2008yv}.
In turn, this contribution represents a 
non-zero value for the  $Z\text{-}Z'$ mixing angle, which is strongly constrained by $Z$-pole observables and low energy constraints~\cite{Erler:2009jh}.

It is important to notice that something similar happens in the standard model where the chiral charges 
of the photon $Q=T_3+Y$ and  are not orthogonal to those of the $Z$ boson $T_3-\sin^2\theta_W Q$, hence, 
one effective kinetic mixing arise by one-loop corrections~\cite{Baulieu:1981ux}.

\section{Benchmark models in $E_6$\label{sec:models}}
\begin{center}
\begin{figure}[h]
\vspace{0.5cm}
 \scalebox{0.9}{
\begin{tabular}{c}
\centerline{\includegraphics[scale=1]{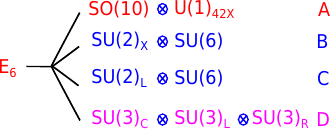}}
\end{tabular}
}
\caption{$E_6$ maximal subgroups}
\label{fig:e6}
\end{figure}
\end{center}

The maximal  subgroups of $E_6$ which can include  
$SU(3)\otimes U(1)$ as an unbroken  symmetry  are~\cite{Slansky:1981yr} $Sp_8$, $SU(2)\otimes SU(6)$, $SO(10)\otimes U(1)$, $F_4$, and $[SU(3)]^3$. 
By imposing the SM  gauge group as an intermediate step in the breaking chain $E_6\rightarrow SM\rightarrow SU(3)\otimes U(1)_{\text{EM}}$, 
the subgroups $Sp_8$  and $F_4$ can be eliminated. So, from now on we are going  to focus only on the breaking chains  in figure~(\ref{fig:e6})  
which,  by  the way, sets part of our convention in the sense  that we refer to  $A$  
as  the chain  belonging to $SO(10)\otimes U(1)$, $B$ to the chain $SU(2)\otimes SU(6)$,$\cdots$, {\em etc}.  

\subsection{$SO(10)\otimes U(1)${}}
\label{sec:so10}
In what follows, the models  will also  be denoted according to the  generalized RR 
notation~\cite{Rojas:2015tqa}. 
The list of models  and their respective RR notations are shown in table~\ref{tab:models1}.

\begin{figure}[h]
\centering 
 \scalebox{1}{
\begin{tabular}{c}
\includegraphics[scale=1]{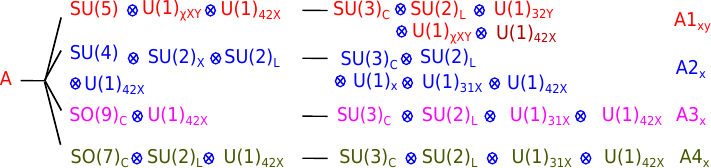} 
\end{tabular}
} 
\caption{$E_6\rightarrow SO(10)\otimes U(1)_{42X}$  chains of subgroups, where $X,Y=R,I,A$ and $X\ne Y$.}
\label{fig:a}
\end{figure}

In $E_6$ there are only three chains of subgroups for which the SM hypercharge $U(1)_Y$~($U_{32I}$ in RR notation) 
appears in  a natural way.  
Two of them, $A1_{RI}$  and $A1_{AI}$~(see  figure~(\ref{fig:a}) and table~\ref{table:a}), go trough 
$SO(10)\otimes U(1)$ and that is one of the reasons why this group have been widely studied in GUTs. 
The $ A1_{RI}$ chain of subgroups corresponds to the embedding of
the Georgi-Glashow unification model~\cite{Georgi:1974sy} $ SU (5) $ in $ E_6$
through the breaking~\cite{Slansky:1981yr} 
$E_6\rightarrow SO(10)\otimes U(1)_{42R}\rightarrow  SU(5)\otimes U(1)_{\chi RI}\otimes U(1)_{42R}\rightarrow
SU(3)\otimes SU(2)\otimes U(1)_{32I}\otimes U(1)_{\chi RI}\otimes U(1)_{42R}$.
The charges of  $U(1)_{\chi RI}$ and $U(1)_{42R}$ corresponds to those of~\cite{Barr:1981qv}  $Z_{\chi}$
and $Z_{\psi}$~(see table~\ref{tab:chiral}), respectively;
these models  are   well known in  $E_6$~(see table~\ref{tab:models1}).
After we rotate to the  mass eigenstate basis  two vector bosons 
  $Z^{\prime}$ and $Z^{\pp}$  appear in addition to the SM fields. 
When the mixing between the SM  $Z$ and the extra neutral vector bosons is zero~\cite{Erler:2009jh,Erler:2009ut,Erler:2010uy,Erler:2011iw} the 
$Z'$ and $Z''$ fields are  a linear combination of~~\cite{London:1986dk}  $Z_{\chi}$  and $Z_{\psi}$
 \begin{align}\label{eq:londonmix}
  Z^{\prime}= \cos \beta Z_{\chi}+\sin \beta Z_{\psi}\ .
 \end{align}

By varying  $\beta$ from 0 to $\pi/2$  the parameter space in figure~(\ref{fig:mixing}) corresponds to the vertical line which goes through $Z_\psi$~($U(1)_{42R}$) 
and $Z_{\chi}$~($U(1)_{\chi RI}$).
That is the parameter space of the models orthogonal to the SM hypercharge, \ie
$\sum_{f\in {\bf 27}}Q_{Z'}(f)Y(f)=0 $, where $Q_{Z'}(f)$ is the $Z'$ charge of the fermion $f$ and $Y(f)$ the  SM hypercharge. 
As was  shown in section~\ref{sec:kmixing}, this vertical line also  corresponds to the parameter space for any   $E_6$-motivated   $Z'$ 
at the unification limit; however, owing to the RGE, at low energies  the values of the $g_{\kappa}$ couplings  will depend on the specific details of the breaking pattern.
Because  at low energies  the couplings  are no longer identical, the $Z'$ parameter space acquires
a component in the SM hypercharge axis in figure~(\ref{fig:mixing}), which is equivalent to a kinetic mixing of the form~\cite{Erler:2011ud}
\begin{align}\label{eq:erlermix}
Z^{\prime} = \cos \alpha \cos \beta Z_{\chi}+\sin \alpha \cos \beta Y+\sin \beta Z_{\psi}\ . 
\end{align}
Due to this mixing  the   $Z'$  parameter space will be out of the unification vertical line as is shown for some models in figure~(\ref{fig:mixing}).

\begin{table}
\tbl{
$E_6$-motivated $Z'$ benchmark models and their  generalized RR  notations~\cite{Rojas:2015tqa}. 
The  $Z_I$, $Z_{\not{d}}$, and  $Z_{\not L}$ bosons are blind, respectively, 
to  up-type quarks,  down-type quarks, and  SM leptons.
Similarly, the $Z_{\not n}$ and the $Z_{\not p}$ are gauge  bosons which do not couple~(at vanishing momentum transfer and at the tree level)
to neutrons and protons, respectively.
 The $Z_{B-L}$ couples purely vector-like while the $Z_\psi$ has only axial-vector couplings to the ordinary fermions. 
For convenience the models with the same multiplet structure as the  $Z_{\chi}$ are referred to as $U_{\chi XY}$. The $Z_S$ model does not have RR notation}
{
\begin{tabular}{|c|c|c|c|c|c|c|}
\hline
$Z'$                                     &
$ Z_R$~\cite{Robinett:1982tq}            & 
$Z_{\not{d}}$~\cite{Erler:2011ud}        & 
$   -  Z_I$~\cite{Robinett:1982tq}       & 
$   -  Z_{L_1}$~\cite{Robinett:1982tq}   & 
$   -  Z_{R_1}$~\cite{Robinett:1982tq}   & 
$      Z_{\not p}$~\cite{Erler:2011ud}      \\ 
RR                & 
$ U_R$            & 
$ U_{A}$          & 
$ U_I $           & 
$ U_{33}$         & 
$ U_{21\ovl{R}}$  & 
$ U_{21\ovl{A}}$  
\\ 
\hline 
\hline     
$Z'$&  
$   -  Z_{\not n}$~\cite{Sanchez:2001ua,Erler:2011ud}& 
$   -  Z_{B-L}$~\cite{Pati:1974yy}   &  
$ Z_{ALR}$~\cite{Ma:1986we}                  & 
$-  Z_{\not L}$~\cite{Babu:1996vt}           & 
$ Z_\psi$~\cite{Robinett:1982tq}             & 
$ Z_\chi$~\cite{Robinett:1982tq}              
\\    
RR                  & 
$ U_{21\ovl{I}}$    & 
$ U_{31R}$          &
$ U_{31A}$          & 
$ U_{31I}$          & 
$ U_{42R}$          & 
$ U_{\chi RI}$       
\\
\hline 
\hline     
$Z'$                                         &  
$ Z_N$~\cite{Ma:1995xk,King:2005jy}          &
$Z_{\chi^{*}}{[\text{flipped-}SU(5)]}$~\cite{Barr:1981qv}&
$ Z_\eta$~\cite{Witten:1985xc}               & 
$ Z_Y$~\cite{Glashow:1961tr,Weinberg:1967tq} & 
$Z_S$~\cite{Erler:2002pr,Kang:2004pp}        &
$ Z_{331G}$~\cite{Singer:1980sw,Sanchez:2001ua,Rodriguez:2016cgr}         
\\    
RR                  & 
$ U_{\chi AI}$      & 
$ U_{\chi RA}$      &
$ U_{51I}$          & 
$  U_{32I}$         & 
                    &
$U_{21\bar{I}}$, $U_{33}$
\\
\hline 
\end{tabular}}
\label{tab:models1}
\end{table}

The other chain of subgroups in which the SM hypercharge appears naturally is $A1_{AI}$.
The $SU(5)$ is  the Georgi-Glashow one, but  the  factor $U(1)_{\chi AI}$ is an alternative version 
of $U_{\chi}$~($U(1)_{\chi RI}$), which is known in the literature as $U_{N}$. 
Figure \ref{fig:a} shows  the   embedding of $SU(5)\otimes U(1)_N$~(\ie $SU(5)\otimes U(1)_{\chi AI} $ ) 
in $E_6$. This is the symmetry group of the Exceptional  Supersymmetric Standard Model~(ESSM)~\cite{King:2005jy},
which is obtained from the  $E_6$ charges   by requiring vanishing  $U(1)_N$ charges for right-handed neutrinos.

Table~\ref{table:a} shows the  six possible ways to embed $SU(5)$ into $SO(10)\otimes U(1) \subset E_6$ (all 
the chains of subgroups of the form $A1_{XY}$ can be seen  in figure~(\ref{fig:a})); from these, the  $A1_{RA}$
chain  corresponds to the flipped $SU(5)$~\cite{Barr:1981qv}. 

Taking as inputs  the values of the fine-structure constant and the  
corresponding quantities for the  strong and weak interactions,
we find the strength couplings at low energies by  using   
the one-loop  RGE equations~\cite{Robinett:1982tq};
 however, we always find  
that for the $A1_{XY}$  chains of subgroups it is not possible 
to get the right order between the unification scales. That problem is related 
to  the wrong prediction of the Weinberg angle in $SU(5)$.
Although it is not possible to have a consistent picture 
for  the embeddings  $SU(5)\otimes U(1)\subset SO(10)\otimes U(1)\subset E_6$, there are  solutions
in most of the remaining  $E_6$ breaking patterns.

\subsubsection{$SU(4)\otimes SU(2)_L\otimes SU(2)\otimes U(1)\subset SO(10)\otimes U(1)\subset E_6${}}
From the three chains of subgroups $A2_{X}$ in figure~(\ref{fig:a}) we can get  
low-energy $E_6$ models~(LEE6Ms)
\ie  models where  at least one of the neutral currents in Eq.~(\ref{EqY2}) does not contribute to the
hypercharge, therefore, the corresponding vector boson is not necessary to have a consistent model. 
Usually, the fermion content of these models is smaller than the fundamental representation of $E_6$.  

\begin{figure}[h]
\centering 
\begin{tabular}{c}
\includegraphics[scale=0.8]{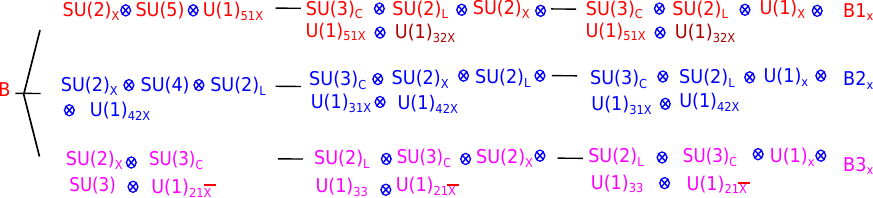} 
\end{tabular}
\caption{$E_6\rightarrow SU(2)_X\otimes SU(6)$  chains of subgroups, where $X=R,I,A$.}
\label{fig:b}
\end{figure}

\begin{figure*}[h]
 \scalebox{0.8}{
\centering 
\begin{tabular}{c}
\includegraphics[scale=1]{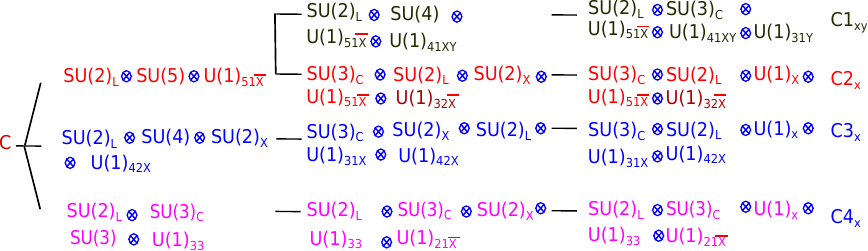} 
\end{tabular}
} 
\caption{$E_6\rightarrow SU(2)_L\otimes SU(6)$  chains of subgroups, 
where $X,Y=R,I,A$ and $X\ne Y$.}
\label{fig:c}
\end{figure*}

The  $A2_{R}/U(1)_{42R}$ chain~\footnote{$A2_{R}/U(1)_{42R}$   denote  the chain of subgroups $A2_{R}$ without the $U(1)_{42R}$ factor.}
is the  Pati-Salam model~\cite{Pati:1974yy,Mohapatra:1974hk} (see figure~(\ref{fig:a})).
The EW charges of this model are the same as those of 
$B2_{R}/U_{42R}$~(see figure~(\ref{fig:b})) and   $C3_R/U_{42R}$~(see figure~(\ref{fig:c}))
and are the same as the Left-Right~(LR) symmetric model. 
The $A2_{A}/U_{42A}$ chain of subgroups  corresponds to  the alternative left-right model~$Z_{ALR}$~\cite{Ma:1986we}. 
The EW charges  for this model were reported in~\cite{Rodriguez:2016cgr} 
and are identical to those of  $B2_{A}/U_{42A}$ and $C3_{A}/U_{42A}$.
$A2_{I}$ is a new model in the literature even though is closely related 
to the second alternative model obtained from trinification~\cite{Rodriguez:2016cgr};
the difference lies in the Abelian factor $U_{42I}$ (in~\cite{Rodriguez:2016cgr}
$Y$ is a linear combination of $U_{31I}$  and $U(1)_I$, while
in the $SO(10)$ embedding  $U_{42I}$  is in place of $U_{I}$).
Identical EW charges are obtained from  $B2_{I}$ and $C3_{I}$.
Note that the coefficients of the hypercharge in   $A3_{I}$ and   $A4_{I}$ are identical to those of
 $A2_I$; however, due to the absence of the $U(1)_{I}$ factor  in the chain of subgroups, 
in  Eq.~(\ref{eq:zprimecharges}) 
there is no mixing with the corresponding vector boson $Z_{I}$. 
In $E_6$ the parameter space   for every  $Z'$  of the $A2_{X}$ chain coincides 
with 	those of $C3_X$ in figure~\ref{fig:mixing}.

\subsection{$SU(2)\otimes SU(6)\subset E_6${}}
The third chain of subgroups in which 
the SM hypercharge $U(1)_{Y}$  appears in a natural way is 
the $B1_{I}$ in figure~(\ref{fig:b}).
This model occurs in Calabi-Yau compactifications in string theory~\cite{Witten:1985xc}
and is commonly denoted as $Z_{\eta}$.
The charges of this model correspond to 
those of $U(1)_{51I}$~(see Table~\ref{tab:chiral}).
In this chain of subgroups the $SU(5)$ is the same as that of   Georgi-Glashow; 
however, the  $U(1)_{51}$ factor is different from  the corresponding factor in  the embedding through $SO(10)$. 
The other two chains  in $SU(2)_{X}\otimes SU(6)$ are    $B2_{X}$ and $B3_{X}$. For these chains,  the $Z'$ 
charges of the LEE6Ms correspond to those of the Pati-Salam and  trinification models, and their corresponding alternative versions, which 
have been studied in the previous section and in reference~\cite{Rodriguez:2016cgr}. 
New models appear in the chains of subgroups containing $SU(2)_L\otimes SU(6)$; of particular interest are $C2_X$,  which contain a $SU(5)$ different 
from the Georgi-Glashow one.  This new $SU(5)$ allows a solution for the mass scales in $E_6$ 
from the one-loop RGE~\cite{Robinett:1982tq} (we saw above that such a solution is not possible either in 
the Georgi-Glashow model or its alternative versions). The same is true for 
the $C1_{xy}$ chains of subgroups. 
The chains  $C3_{X}/U(1)_{42X}$ and $C4_X$  
have the same  $Z'$ charges as the Pati-Salam and trinification models, respectively. 
The low-energy  $Z'$ charges for $C1_X$, $C2_X$ and $C3_X$ as a function of the $\theta$ mixing angle~(see  Eq.~(\ref{eq:zprimecharges})) 
are shown in the Sanson-Flamsteed projection in figure~(\ref{fig:mixing}).  

\subsection{$SU(3)\otimes SU(3)\otimes SU(3)${}}
The $SU(3)\otimes SU(3)\otimes SU(3)$ is the gauge group 
of trinification~\cite{Achiman:1977py,Achiman:1978vg,Glashow:1984gc,Babu:1985gi,Bai:2017zhj}. 
In table~\ref{table:d} are shown the three possible chains of breakings for this group.
As was shown in reference~\cite{Rodriguez:2016cgr},
the charges of the three chains reduce 
to those of the universal 331 vector boson  $Z_{331G}$ ~(see table~\ref{tab:e6limits} for the LHC constraints).  
A detailed study of these models and their EW 
constraints was presented in reference~\cite{Rodriguez:2016cgr}.

\section{Low-energy models without unification.}
\label{sec:lem}
\begin{figure}[h]
\centering 
\begin{tabular}{cc}
\includegraphics[scale=0.2]{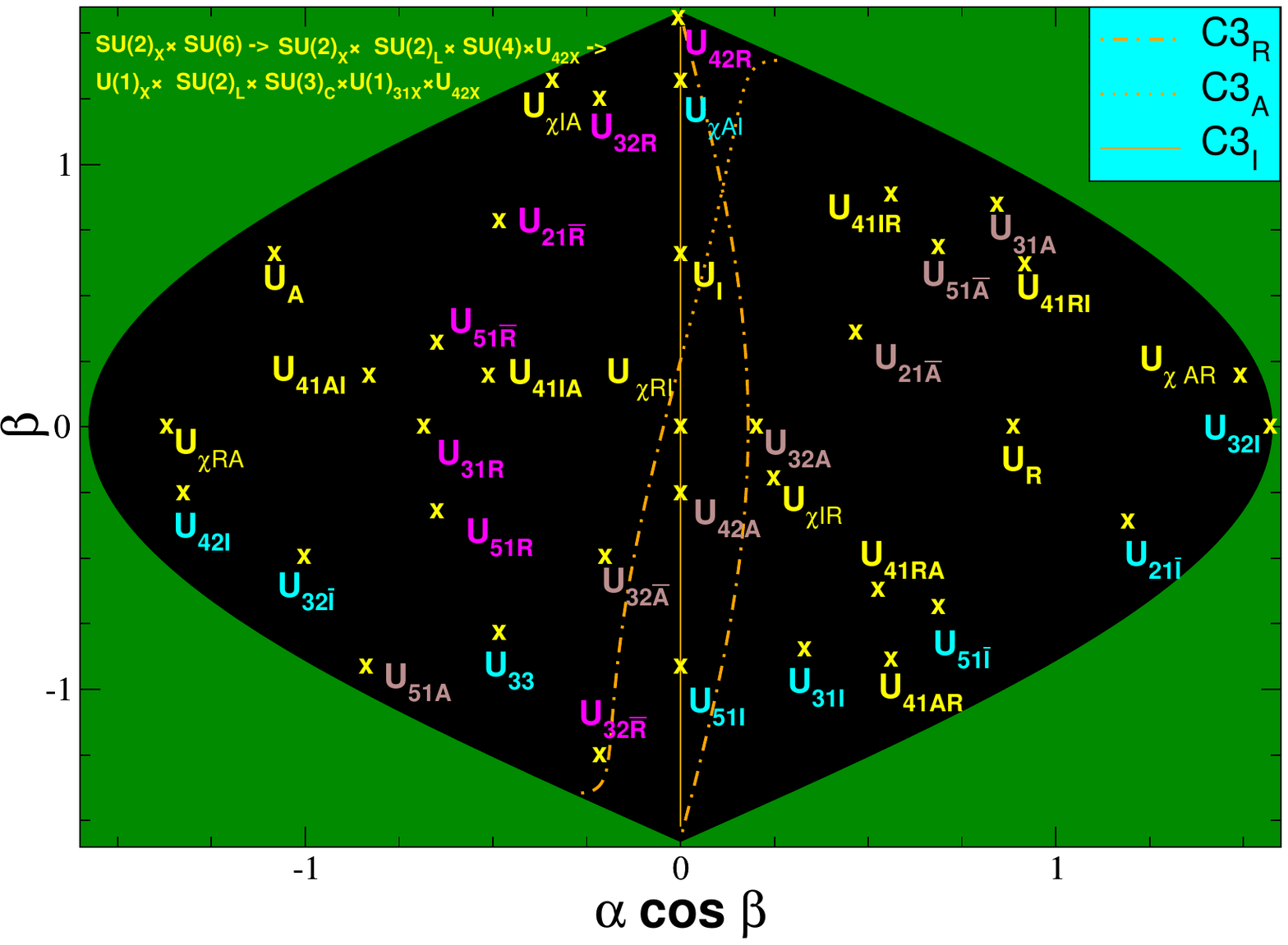}   & \includegraphics[scale=0.2]{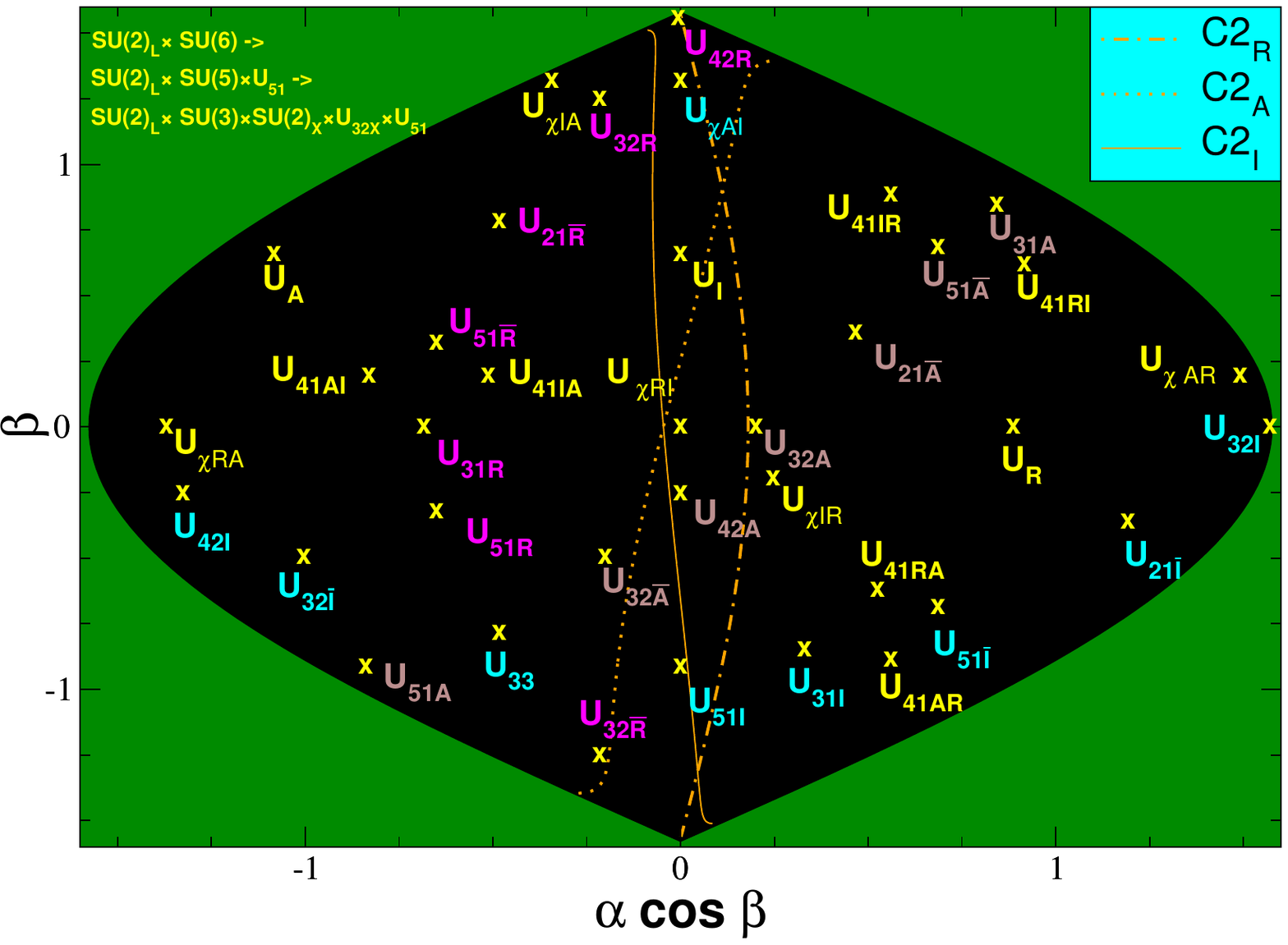} \vspace{-12pt} \\
\includegraphics[scale=0.2]{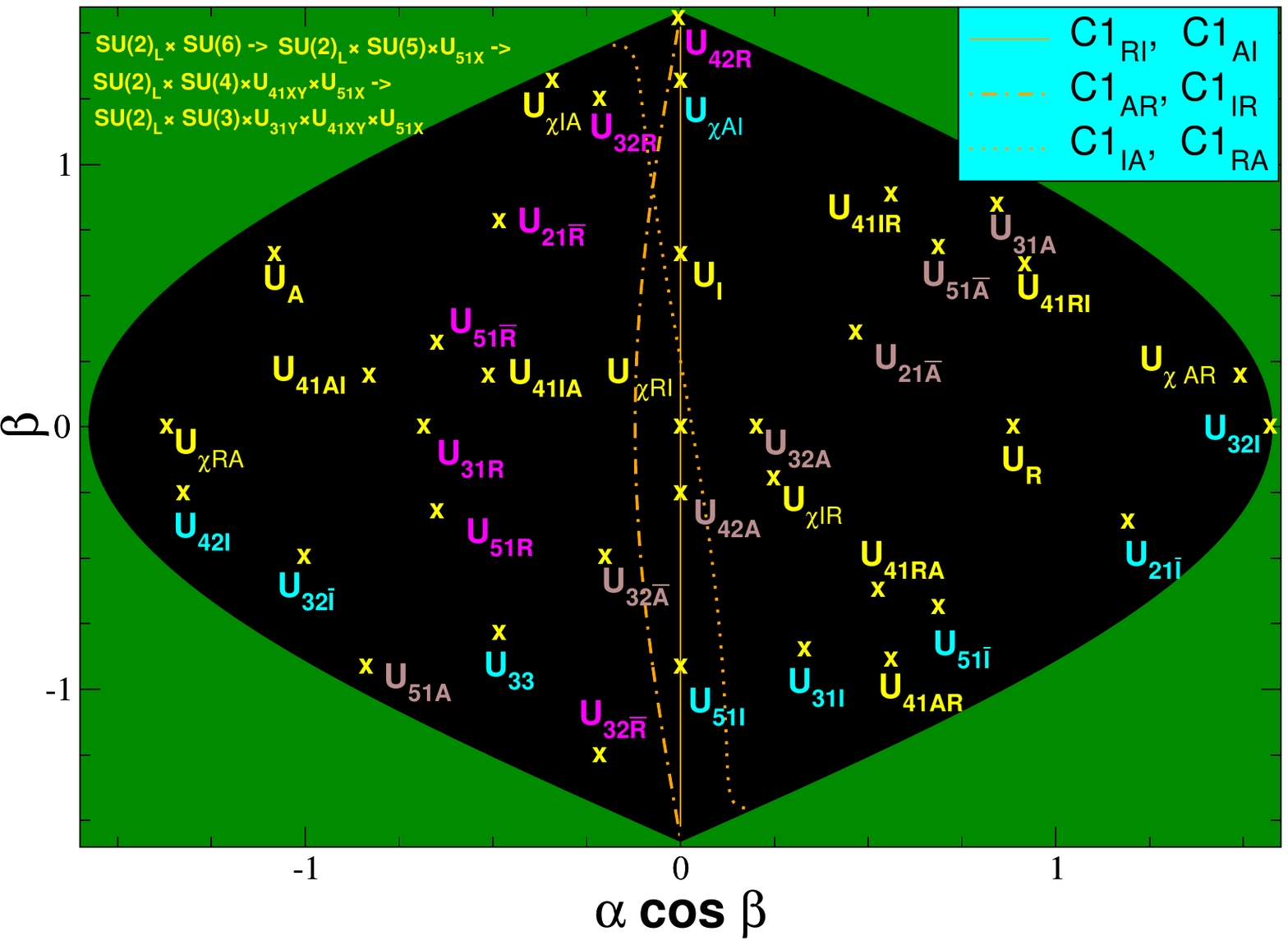}   & \includegraphics[scale=0.2]{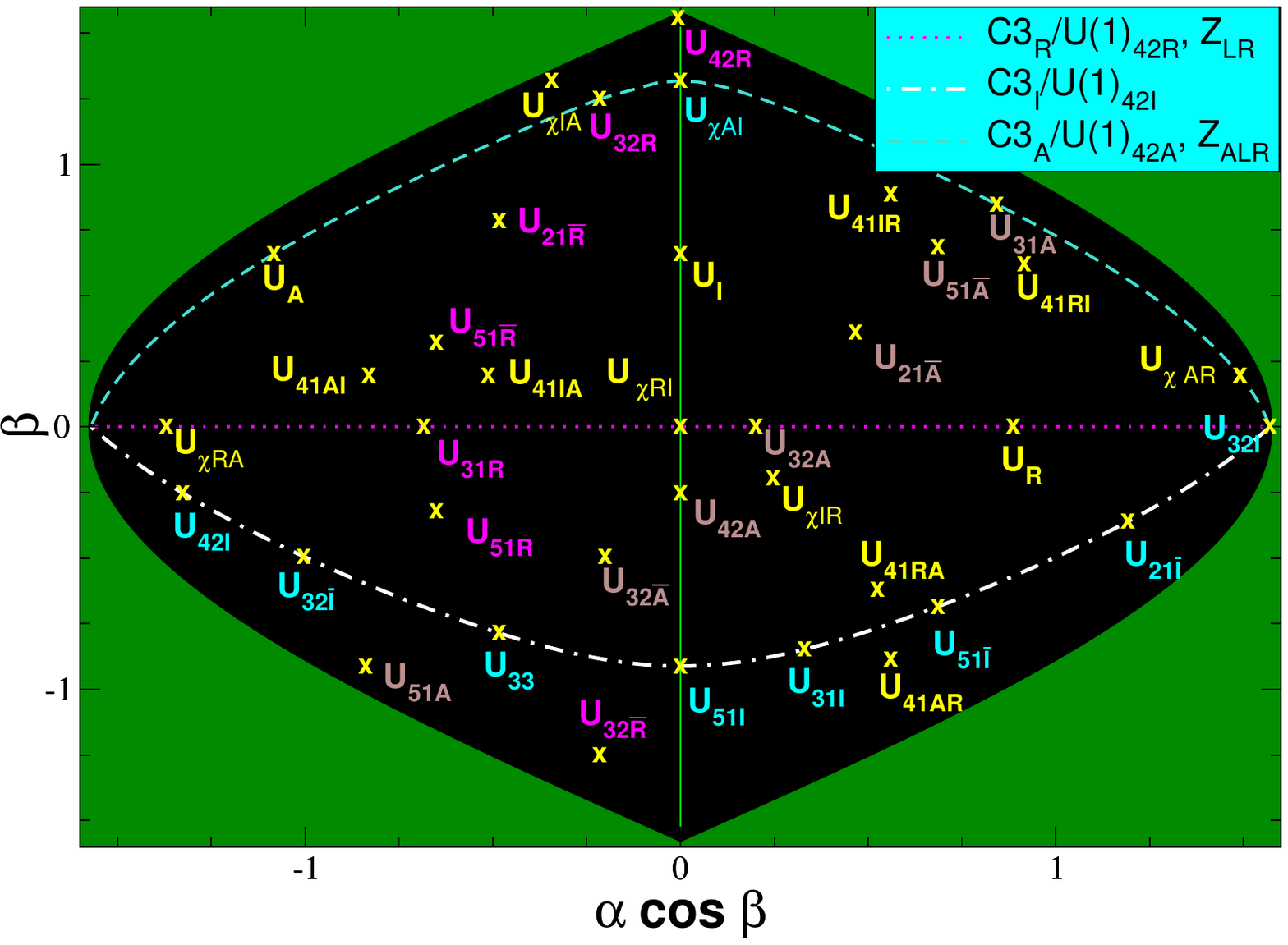} \\
\end{tabular}
\caption{Sanson-Flamsteed projection of the $\alpha$-$\beta$    parameter space in $E_6$. 
The vertical line  $\cos\beta = 0$  represent the models with charges orthogonal to the SM hypercharge, thus, deviations of this line 
show that at low energies there is a mixing between the $Z'$ and the field associated with the SM hypercharge.
To obtain the values of $\alpha$ and $\beta$ 
in the top figures and in the bottom-left one
we used the RGE~\cite{Robinett:1982tq} in order to get the $g_{a}$, $g_{b}$  and $g_{c}$ coupling strengths  at the EW scale 
(from their initial values at the grand unification scale),
then,  we solved  Eq.~(\ref{eq:charges})  varying  $\theta$ between $-\pi$ and $\pi$. 
For the bottom-right figure, we put aside the unification hypothesis and, by ignoring the mixing with the fields associated
with the charges that do not contribute to the electric charge,
we explored the possible values for the coupling strengths; see section~\ref{sec:lem} for additional details.}
\label{fig:mixing}
\end{figure}

The most general charges  of any  $E_6$-motivated 
$Z'$ model is generated by  the linear combination of three  independent 
sets of charges associated with the $U(1)_\kappa$ symmetries, where $\kappa=a,b,c$.
In appendix~\ref{sec:unification}, we showed that for unification models the values of $\alpha$ 
and $\beta$ corresponds to the vertical line which passes through $Z_\psi$ and $Z_\chi$. 
At low energies the parameter space of
these models keeps close to this  line, as can be seen 
in figure~(\ref{fig:mixing}).

Without the unification hypothesis it is not possible
to determine the preferred region in the parameter space. 
There are several models based on $E_6$ subgroups, and  in some of them
unification is not necessary to get a  predictable model.
In most of the well-known  cases, the subgroup rank is less than the  $E_6$ rank
and at least one of the vector currents does not contribute to the electric charge. 
In order to ignore this current  we set $\theta=0$ in Eq.~(\ref{eq:zprimecharges}), in such a way that for a fixed value of the couplings the $Z'$ 
charges reduce to a single point in the Sanson-Flamsteed projection. 
Since the values of the couplings is arbitrary, 
by varying them  we generate the parameter space for these models.

For these  models the hypercharge  
is the combination of the charges of two $U(1)'\text{s}$. If we put    $k_c= 0$  and $\sinx= 0$, the charges in Eq.~(\ref{eq:zprimecharges}) reduce to 
\begin{align}\label{eq:chiral}
\gzp\epsilon^{Z^\prime}_{L,R}=& \gp \left(k_a \frac{\epsilon^{b}_{L,R}}{\hb}-\hb  k_b \epsilon^{a}_{L,R}   \right)\ .
 \end{align}
Owing to the fact that  $Q_c$ does not contribute to the electric charge, 
in these models is possible to have a low-energy theory without 
the corresponding   $Z^{\prime\prime}$ associated with $U(1)_c$. 
In section~\ref{sec:models} we denoted them as  LEE6Ms. 
In $E_6$ there are three chains of subgroups  where one of the $Q_{k}$ corresponds 
to the SM hypercharge, in these cases, the SM is the LEE6M and, in principle, it
does not require from other vector bosons to be a consistent theory. 
In  Eq.~(\ref{eq:chiral})  the  $U_a$ and $U_b$ charges  appear in a symmetric way, except
by a global sign which, in general, can not be determined from the symmetry group.

In panel~(\ref{fig:mixing}) the  bottom-right figure shows 
the parameter space of some models based on $E_6$ subgroups. 
The horizontal dotted magenta line corresponds to the parameter space of the well-known LR models, 
which are LEE6Ms in the chains of subgroups  $A2_R$, $B2_R$ and the $C3_R$. 
As expected, in this line appears the charges of the $Z_{B-L}$~($U_{31R}$) and $Z_R$~($U_R$). 
This line also represents the set of possible $Z'$ models for flipped $SU(5)$ which are a 
linear combination of the $U_{\chi RA}$
and $U_{32A}$. 
The dashed cyan line contains the   $Z'$ parameter space
 of the alternative left-right model $Z_{ALR}$.
These models are the linear combination of $U_{31A}$
and $U_A$~(the downphobic model $Z_{\not{d}}$). 
This line also corresponds to the possible $Z'$ of 
the LEE6M of the chain of subgroups $A1_{AR}$ which has not been reported in the literature, as far as we know.
The dot-dashed gray line is the set of the possible $Z'$ models of the LEE6M associated 
with the chain of subgroups $C4_I$ which contains the universal 331 model~\cite{Singer:1980sw,Sanchez:2001ua,Rodriguez:2016cgr}. 
We obtain these models from the linear combination of the $U(1)_{33}$ and the $U_{21\bar{I}}$, which have the quantum numbers 
of $\lambda_{8L}$  and  $U(1)^{\prime}$ in the 331 models, respectively. This line is also generated from the third alternative 
left right model~\cite{Rodriguez:2016cgr} and results from the linear combination of $U_{31I}$ 
(the leptophobic model $Z_{\not{L}}$) and  $U_{42I}$. This line also
corresponds to the possible $Z'$ for the  LEE6M of $C2_{I}$, 
which, to the best of our knowledge, has not been reported in the literature.  
This set of points contains the $Z_{\eta}$~($U_{51I}$) model.
\section{LHC constraints}
\label{sec:constraints}
\begin{figure*}[h]
\centering 
\begin{tabular}{cc}
\includegraphics[scale=0.2]{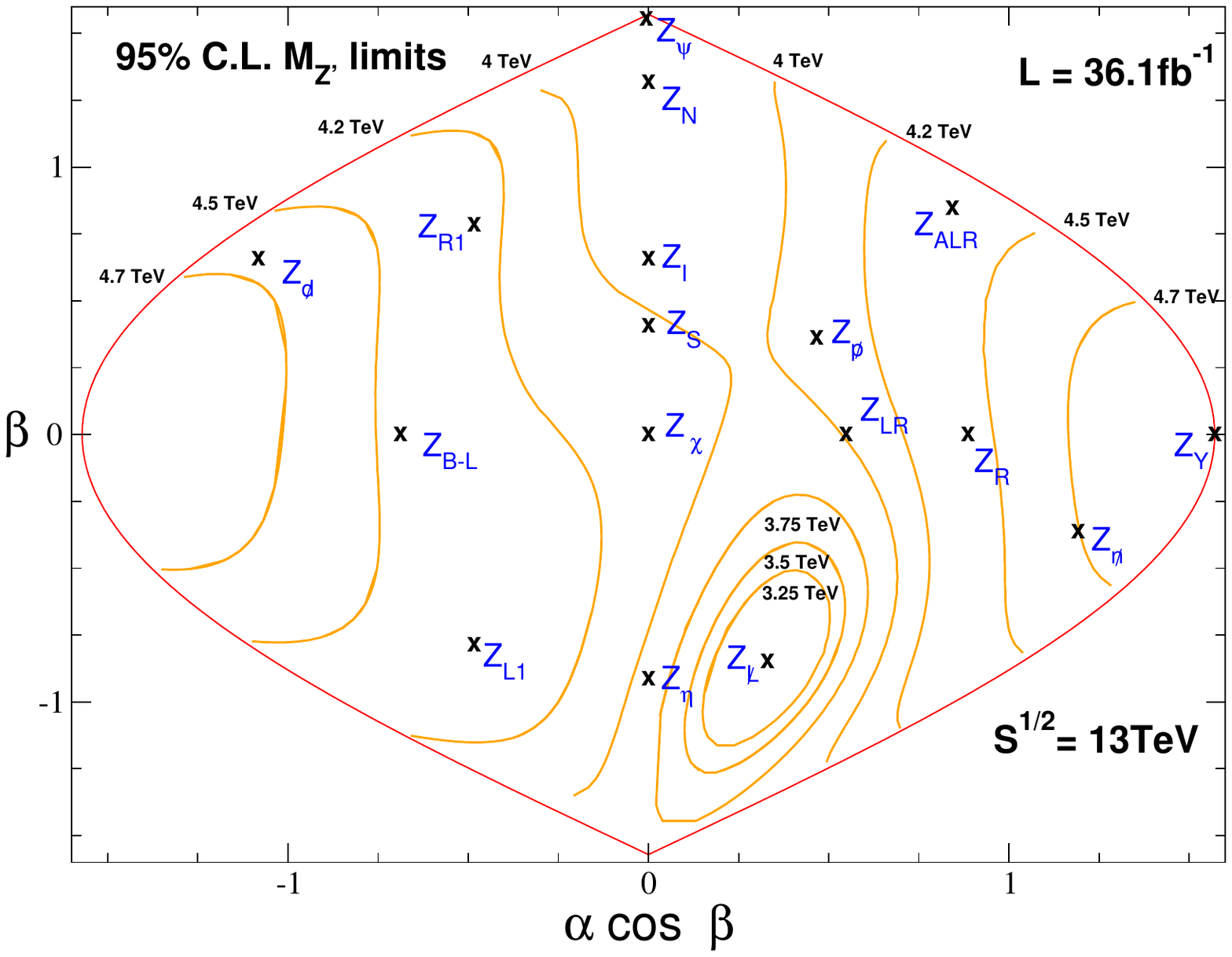} & \includegraphics[scale=0.2]{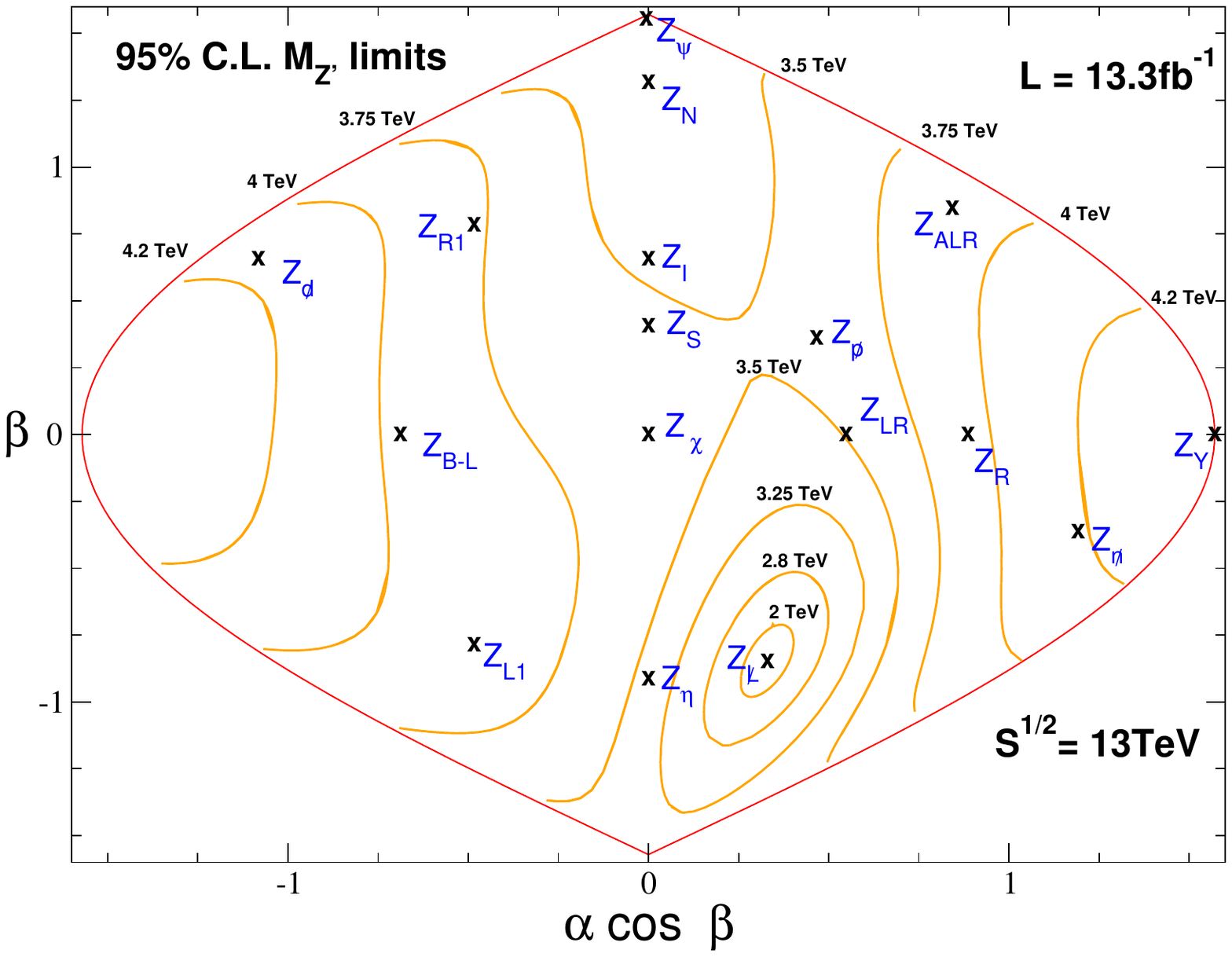} \\
\end{tabular}
\caption{$\alpha$-$\beta$ Sanson-Flamsteed projection of the $Z'$  parameter space in $E_6$. 
The contours show the 95\% C.L. lower limits  on the $Z^{\prime}$ mass~(in TeV).}
\label{Contours}
\end{figure*}

Finally, we  also  report  the most recent constraints 
from colliders and low-energy experiments on the neutral current  parameters  for  some $Z'$-motivated $E_6$ models and the sequential standard model~(SSM).
For the time being, the strongest constraints  come  from the   proton-proton collisions  data collected by the ATLAS experiment 
at the LHC  with an integrated luminosities  of 36.1~fb$^{-1}$ and 13.3~fb$^{-1}$  
at a  center of mass energy of 13 TeV~\cite{Aaboud:2017buh,ATLAS:2016cyf}.  In particular, 
we used the upper limits  at 95\% C.L.  on the total cross-section of the $Z'$ decaying
into dileptons~(\ie  $e^+e^-$  and  $\mu^+\mu^-$).
Figure~(\ref{Contours}) shows the contours of the lower limits on the $M_{Z^{\prime}}$ at 95\% C.L.
We obtain these limits from the intersection of  $\sigma^{\text{NLO}}(pp\rightarrow Z'\rightarrow l^{-}l^{+})$
with the ATLAS 95\% C.L. upper limits on the cross-section~(for additional details see  reference~\cite{Salazar:2015gxa}).
As a cross-check we calculated these limits for some models as shown in table~\ref{tab:e6limits} for various  $E_6$-motivated $Z'$ models and the SSM model.
In order to compare, we also show in this figure  the constraints for   all the models reported by ATLAS. 
For the 36.1~fb$^{-1}$ data we multiply the theoretical cross-section by a global $K$ factor 
to reproduce the ATLAS  constraints for the $Z_{\chi}$ model.  
This procedure was not necessary for the 13.3~fb$^{-1}$ dataset.

\begin{table}[ht]
\tbl{95$\%$ C.L. lower mass  limits~(in TeV) for $E_6$-motivated $Z'$ models and the sequential standard model~$Z_{\text{ssm}}$. 
These constraints come from the 36.1~fb$^{-1}$ and  13.3~fb$^{-1}$ datasets for proton-proton collision 
 at a center of mass energy of $\sqrt{s}=13$~TeV~\cite{Aaboud:2017buh,ATLAS:2016cyf}.
}{
\noindent\makebox[\textwidth]{
\scalebox{0.8}{
\begin{tabular}{|l|l|r|r|r|r|r|r|r|r|r|r|r|r|}
\hline
 $Z^{\prime}$ model &luminosity&
 $Z_\chi$ &
 $Z_\psi$ &
 $Z_\eta$&
$Z_{LR}$&
$Z_{R}$&
$Z_{N}$&
$Z_{S}$&
$Z_{I}$&
$Z_{B-L}$&
$Z_{\not{d}}$&
$Z_{331G}$&
$Z_{SSM}$
\\
\hline
$M_{Z'}$($^\dagger$fitted)& (36.1fb$^{-1}$) &
4.1$^{\dagger}$&
3.81&
3.91&
4.28&
4.41&
3.84&
4.02&
3.94&
4.44&
4.66&
4.608    &
4.58\\
\hline
ATLAS& (36.1fb$^{-1}$) &
4.1&
3.8&
3.9&
---&
---&
3.8&
4.0&
4.0&
---  &
---  &
---  &
4.5\\
\hline
\hline
$M_{Z'}$& (13.3fb$^{-1}$) &
3.62&
3.35&
3.43&
3.77&
3.92&
3.38&
3.54&
3.47&
3.95&
4.15&
4.10&
4.05\\
\hline
ATLAS& (13.3fb$^{-1}$) &
3.66&
3.36&
3.43&
---&
---&
3.41&
3.62&
3.55&
---  &
---  &
---  &
4.05\\
\hline
\end{tabular} }}
\label{tab:e6limits}
}
\end{table}

\section{Conclusions}
\label{sec:conclusions}
In the present work 
we have reported the most general expression for the chiral charges of a 
neutral gauge boson $Z'$ coming from an $E_6$ unification model,
in terms of the EW parameters  and  the charges of
the $U(1)$ factors in the chain of subgroups.

 We also showed for any breaking pattern that, 
 the charges of 
SM hypercharge are orthogonal to the corresponding charges 
of the  $Z'$ gauge boson~\ie  $\sum_{f\in 27} Y(f)Q_{Z'}(f)=0$ (see section~\ref{sec:kmixing}),
if the 
values of the  $g_{\kappa}$ coupling strengths  associated with the $U(1)_{\kappa}$ 
factors of the chains of breakings are equal to each other. 
Due to the RGE  the couplings are no longer identical at low energies, therefore  
there must be a mixing between the field associated with $Y$ and the  $Z'$.   
This mixing can modify several observables as it has been shown in  reference~\cite{Holdom:1985ag,Babu:1997st,Langacker:2008yv}.

Pure neutral gauge bosons coming from $E_6$  are $Z_\psi$ and $Z_\chi$
as introduced  in section~\ref{sec:so10} but the physical neutral states $Z'$ and $Z''$
are a mixing of those states according to Eqs.~(\ref{eq:londonmix}) and (\ref{eq:erlermix}),
which define the  $\alpha$ and $\beta$  angles in our analysis. 
By using the RGE~\cite{Robinett:1982tq} and assuming $E_6$ unification, we showed that 
for most of the chains of breaking in $E_6$ it is possible to solve 
the equations for the mass scales 
in a consistent way~(one important exception are the chains 
of subgroups that contain the Georgi-Glashow $SU(5)$ model  and their alternative versions). 
This procedure allowed us to calculate 
the low-energy coupling strengths for several chains
of subgroups and the $Z'$  parameter space in the Sanson-Flamsteed projection.
It is worth noting that in  $E_6$ unification  at low energies the parameter space of
these models keeps close to the mentioned vertical line 
as can be seen in  figure~(\ref{fig:mixing}). 
To the best of our knowledge, several of the analyzed chains of subgroups  presented here are new in
the literature.

The most general charges  of any  $E_6$-motivated 
$Z'$ model is generated by  the linear combination of three  independent 
set of charges associated with the different  $U(1)$ symmetries.
By putting aside the unification hypothesis it is not possible
to determine the preferred region in the parameter space; 
however, by ignoring the mixing with the
associated charges that do not contribute to the electric charge, 
the corresponding parameter space reduces to a single line in the $\alpha$-$\beta$ Sanson-Flamsteed
projection as shown for some models in the bottom right figure in~(\ref{fig:mixing}). 

By using the most recent upper limits  on the cross-section 
for extra gauge vector bosons  $Z'$ decaying into dileptons
form   ATLAS data at 13~TeV 
with  accumulated luminosities of  36.1fb$^{-1}$~\cite{Aaboud:2017buh} 
and 13.3fb$^{-1}$~\cite{ATLAS:2016cyf} 
for  the Drell-Yang processes $pp\rightarrow Z (\gamma) \rightarrow l^{+}l^{-} $,  
we set 95$\%$ C.L. lower limits  on    the  $Z'$ mass for the typical $E_6$ benchmark models.  
We also reported the contours in the 95\% C.L.  $Z'$ mass limits for the entire $Z'$ parameter space in $E_6$.
Our results are in agreement with the    lower  mass limits  reported by ATLAS   for the $E_6$-motivated $Z'$ models  and the sequential standard model $Z_{\text{SSM}}$. 

Finally it is important to stress that 
the recent  LHCb anomalies could also 
be  explained by $E_6$ subgroups~\cite{Hati:2015awg,Joglekar:2016yap,Das:2016vkr,Dorsner:2017ufx,Blanke:2018sro,Popov:2016fzr}.
A natural continuation of our work would be to find  
which of these  models are able to explain the anomalies.
That is an interesting question since the $E_6$ models, 
in general, have been considered as phenomenologically safe. 

\section*{Acknowledgments} 
R. H. B. and L. M. thank the  ``Centro de Investigaciones  ITM''. 
We thank Financial support from ``Patrimonio Aut\'onomo Fondo Nacional de Financiamiento para la Ciencia, la Tecnolog\'ia
y la Innovaci\'on, Francisco Jos\'e de Caldas'', 
and ``Sostenibilidad-UDEA''.

\appendix

\section{Rotation matrix}
\label{sec:rotation}
Let us now consider an explicit representation for the orthogonal matrix $\mathcal{O}$ in terms 
of three angles $\omega$, $\phi$ and $\theta$, which are allowed to take values in the $[-\pi,\pi)$ interval. For convenience we choose
\begin{align}\label{Eq7}
\mathcal{O} &=
\begin{pmatrix}
1 &0  &0  &0 \\
0 &\cosz &-\sinz &0 \\
0 &\sinz &\cosz &0 \\
0 &0 &0 &1
\end{pmatrix}
\begin{pmatrix}
1 &0  &0  &0 \\
0 &\cosy &0 &-\siny \\
0 &0 &1 &0 \\
0 &\siny &0 &\cosy
\end{pmatrix}
\begin{pmatrix}
1 &0  &0  &0 \\
0 &1 &0 &0 \\
0 &0 &\cosx &-\sinx \\
0 &0 &\sinx &\cosx
\end{pmatrix}\ .\notag 
\end{align}
Relations in Eq.(\ref{Eq4}) imply then that 
\begin{align}
\cosy\cosz = \frac{k_a\gp}{g_a} \ , \hspace{0.5cm} \cosy\sinz = \frac{k_b\gp}{g_b} \ , \hspace{0.5cm} \siny = \frac{k_c\gp}{g_c}\ ,
\end{align}
which allows us to write the $\phi$ and $\omega$ angles in terms of the $g_a$ and $g_b$ coupling 
constants and of the $k_a$, $k_b$ and $k_c$ coefficients. The $\theta$ angle, however, 
cannot be fixed and must be considered to be another free-parameter. 
It is easy to show that 
\begin{align}
 \cosy &= \ha\frac{\gp}{g_c} , \label{cy}\\    
 \cosz &= \frac{k_a g_c}{\ha g_a}\ ,\label{cz}\\
 \sinz &= \frac{k_b g_c}{\ha g_b}\ ,\label{sz} 
\end{align}
where  
\begin{align}
\ha
=g_c\sqrt{\frac{k_a^2}{g_a^2}+\frac{k_b^2}{g_b^2}}
= \sqrt{\frac{g_c^2}{g^2}\cot^2 \theta_W-k_c^2}\ .
\end{align}

\section{$Z^{\prime}${} charges}
\label{sec:zprimecharges}
From the Lagrangian 
\begin{align}
-\mathcal{L}_{NC}=&gJ_{3L}^{\mu}A_{3L\mu}+  g_{a}J_{a\mu}A^{\mu}_{a}
+ g_{b}J_{b\mu}A^{\mu}_{b}+ g_{c}J_{c\mu}A^{\mu}_{c}\notag\\
=&
g_{\kappa}J_{\kappa\mu}A^{\mu}_{\kappa}=
g_{\kappa^{\pp}}
J_{\kappa^{\pp}\mu}
\mathcal{O}_{\kappa^{\pp}\kappa^\prime}
\mathcal{O}_{\kappa^\prime\kappa}^{T}
A^{\mu}_{\kappa}\notag\\
\equiv & g_{\kappa^\prime}J_{\kappa^{\prime}\mu}Z_{\kappa^{\prime}}^{\mu}\ ,
\end{align}
where $\kappa$ runs over $3L,a,b,c$ and $k^{\prime}$ and $k^{\prime\prime}$ runs over $3L,Y,Z^{\prime},Z^{\pp}$, 
from this equation we obtain
\begin{align}
J_{\kappa^{\prime}\mu}=J_{\kappa\mu}\mathcal{O}_{\kappa\kappa^\prime}=\mathcal{O}_{\kappa^\prime\kappa}^T J_{\kappa\mu}\ .
\end{align}
From this equation the current associated with the $Z'$
is given by\footnote{We have omitted a global sign which cannot be determined from the symmetry group.}
\begin{align}
g_{Z^\prime}J_{Z^\prime}=&
\gp \left(-\frac{k_a J_a^{\mu}+k_b J_b^{\mu}}{\ha}k_c+\ha J_c^{\mu} \right)\sinx+
 \frac{g_c}{\ha}\left(\frac{k_aJ_b^{\mu}}{\hb}-\hb k_b J_a^{\mu} \right)\cosx\ ,\\
\gzpp J_{\zpp}=&
\gp \left(-\frac{k_a J_a^{\mu}+k_b J_b^{\mu}}{\ha}k_c+\ha J_c^{\mu} \right)\cosx-
 \frac{g_c}{\ha}\left(\frac{k_aJ_b^{\mu}}{\hb}-\hb k_b J_a^{\mu} \right)\sinx\ ,  
 \end{align}
where $\hb = \frac{g_a}{g_b}$.
For the LEE6Ms~($c=0$  and  $\sinx = 0$) we obtain
 \begin{align}
 \gzp J_{\gzp}^{\mu}=& \gp \left(k_a \frac{J_{b}^{\mu}}{\hb}-\hb  k_b J_{a}^{\mu} \right) 
                    =  \gp \left(k_b \frac{J_{b}^{\mu}}{\haa}k_a-\haa  J_{a}^{\mu}   \right)\text{sign}(k_b)\ , \\
                    =& \gp \left(\hab  J_{b}^{\mu}- k_a\frac{ J_{a}^{\mu}}{\hab}k_b \right)\text{sign}(k_a)\ ,
\end{align}
where 
\begin{align}
\hat{\alpha}_{a,b} = \sqrt{\frac{g_{a,b}^2}{g^2}\cot^2 \theta_W-k_{a,b}^2}\ .
\end{align}

\section{Sanson-Flamsteed Projection}
\label{sec:unification}

As we mentioned in section~\ref{sec:models} in general any $Z'$ in $E_6$ can be written as a linear combination of three linear independent models. 
One usual basis is given by 
\begin{align}\label{eq:charges}
g_{Z^\prime}\epsilon^{Z^\prime}_{L,R}
=&g_{Z^\prime}\left(\cos \alpha \cos \beta \epsilon^{Z_{\chi}}_{L,R}+\sin \alpha \cos \beta Y_{L,R}+\sin \beta \epsilon^{Z_{\psi}}_{L,R} 
  \right)\eta\ ,  \notag\\
=& A_{L,R}\sinx +B_{L,R}\cosx\ , \hspace{0.2cm}\text{with}\hspace{0.2cm}\eta=\pm 1\ ,  
\end{align}
where the $\epsilon^{Z_{\chi}}_{L,R}$  and $\epsilon^{Z_{\psi}}_{L,R}$ are the  chiral charges of the $Z_{\chi}$ and $Z_\psi$ models, respectively.
In this equation  $g_{Z^\prime}$ is given by the Eq.~\ref{eq:gzp}.
In the last line we equate the chiral charges for the $Z'$ associated to a given chain of subgroups Eq.~(\ref{eq:zprimecharges})
to the general expression of the  $E_6$ motivated $Z'$ charges in the $\alpha\text{-}\beta$ parameter space~Eq.~\ref{eq:erlermix}.
We can obtain the partial unification mass scales for every breaking pattern according to 
the reference~\cite{Robinett:1982tq}\footnote{For some breakings there is some ambiguity, in these cases, we chose the lowest mass scale at its minimum value}. 
By evolving   $g_a$, $g_b$ and $g_c$  down to low energies for  every   $\theta$ there is a pair $(\alpha,\beta)$ according with the equation~(\ref{eq:charges}).
$\theta$ parametrizes the mixing between the $Z'$ and $Z''$ the charges~(\ref{eq:charges})  and the corresponding parameter space to low energies 
is shown in figure~(\ref{fig:mixing}).
It is important to notice that at low energies the charges keep close to the vertical line which corresponds to the unification parameter space. 

\section{tables}
\label{sec:tables}

\begin{table}[h]
\tbl{$E_6$ normalized chiral  charges  of ordinary fermions and right-handed neutrinos. Here $l_L = (\nu_L, e_L)^T$ and $q_L = (u_L, d_L)^T$ denote the left-handed lepton and quark doublets.}
{
\scriptsize
\bgroup                    
\def\arraystretch{1.5}
\scalebox{0.96}{
\begin{tabular}{|c|cccccc|}
\hline
\hline  
Model   &$\epsilon_L(l)$  &$\epsilon_R(\nu)$  &$\epsilon_R(e)$  &$\epsilon_L(q)$ &$\epsilon_R(u)$ &$\epsilon(d)$ \\    
\hline
\hline
$U_Y$   &$-\frac{1}{2}\sqrt{\frac{3}{5}}$ &$0$ &$-\sqrt{\frac{3}{5}}$  &$\frac{1}{2\sqrt{15}}$  &$\frac{2}{\sqrt{15}}$  &$-\frac{1}{\sqrt{15}}$\\
\hline
$U_{R}$ &$0$ &$\frac{1}{2}$ &$-\frac{1}{2}$  &$0$  &$\frac{1}{2}$  &$-\frac{1}{2}$\\
$U_{I}$ &$\frac{1}{2}$ &$\frac{1}{2}$ &$0$  &$0$  &$0$  &$-\frac{1}{2}$\\
$U_{A}$ &$\frac{1}{2}$ &$0$ &$\frac{1}{2}$  &$0$  &$-\frac{1}{2}$  &$0$\\
\hline
$U_{33}$ &$\frac{1}{2\sqrt{3}}$  &$\frac{1}{\sqrt{3}}$  &$\frac{1}{\sqrt{3}}$ &$-\frac{1}{2\sqrt{3}}$ &$0$    &$0$\\
\hline
$U_{21\overline{R}}$ &$\frac{1}{\sqrt{3}}$  &$\frac{1}{2\sqrt{3}}$  &$\frac{1}{2\sqrt{3}}$  &$0$  &$-\frac{1}{2\sqrt{3}}$  &$-\frac{1}{2\sqrt{3}}$\\
$U_{21\overline{I}}$ &$-\frac{1}{2\sqrt{3}}$  &$\frac{1}{2\sqrt{3}}$  &$-\frac{1}{\sqrt{3}}$  &$0$  &$\frac{1}{\sqrt{3}}$  &$-\frac{1}{2\sqrt{3}}$\\
$U_{21\overline{A}}$ &$\frac{1}{2\sqrt{3}}$  &$\frac{1}{\sqrt{3}}$  &$-\frac{1}{2\sqrt{3}}$  &$0$  &$\frac{1}{2\sqrt{3}}$  &$-\frac{1}{\sqrt{3}}$\\
\hline
$U_{31R}$ &$\frac{1}{2}\sqrt{\frac{3}{2}}$ &$\frac{1}{2}\sqrt{\frac{3}{2}}$ &$\frac{1}{2}\sqrt{\frac{3}{2}}$  &$-\frac{1}{2\sqrt{6}}$  &$-\frac{1}{2\sqrt{6}}$  &$-\frac{1}{2\sqrt{6}}$\\
$U_{31I}$ &$0$ &$\frac{1}{2}\sqrt{\frac{3}{2}}$ &$0$  &$-\frac{1}{2\sqrt{6}}$  &$\frac{1}{\sqrt{6}}$  &$-\frac{1}{2\sqrt{6}}$\\
$U_{31A}$ &$0$ &$0$ &$-\frac{1}{2}\sqrt{\frac{3}{2}}$  &$\frac{1}{2\sqrt{6}}$  &$\frac{1}{2\sqrt{6}}$  &$-\frac{1}{\sqrt{6}}$\\
\hline
$U_{42R}$ &$\frac{1}{2\sqrt{6}}$  &$-\frac{1}{2\sqrt{6}}$  &$-\frac{1}{2\sqrt{6}}$  &$\frac{1}{2\sqrt{6}}$  &$-\frac{1}{2\sqrt{6}}$  &$-\frac{1}{2\sqrt{6}}$\\
$U_{42I}$ &$\frac{1}{\sqrt{6}}$  &$\frac{1}{2\sqrt{6}}$  &$\sqrt{\frac{2}{3}}$  &$-\frac{1}{2\sqrt{6}}$  &$-\frac{1}{\sqrt{6}}$  &$\frac{1}{2\sqrt{6}}$\\
$U_{42A}$ &$\frac{1}{\sqrt{6}}$  &$\sqrt{\frac{2}{3}}$  &$\frac{1}{2\sqrt{6}}$  &$-\frac{1}{2\sqrt{6}}$  &$\frac{1}{2\sqrt{6}}$  &$-\frac{1}{\sqrt{6}}$\\
\hline
$U_{32R}$ &$\frac{1}{2}\sqrt{\frac{3}{5}}$ &$0$ &$0$  &$\frac{1}{2\sqrt{15}}$  &$-\frac{1}{\sqrt{15}}$  &$-\frac{1}{\sqrt{15}}$\\
$U_{32I}$ &$-\frac{1}{2}\sqrt{\frac{3}{5}}$ &$0$ &$-\sqrt{\frac{3}{5}}$  &$\frac{1}{2\sqrt{15}}$  &$\frac{2}{\sqrt{15}}$  &$-\frac{1}{\sqrt{15}}$\\
$U_{32A}$ &$\frac{1}{2}\sqrt{\frac{3}{5}}$ &$\sqrt{\frac{3}{5}}$ &$0$  &$-\frac{1}{2\sqrt{15}}$  &$\frac{1}{\sqrt{15}}$  &$-\frac{2}{\sqrt{15}}$\\
\hline
$U_{32\overline{R}}$ &$0$ &$\frac{1}{2}\sqrt{\frac{3}{5}}$ &$\frac{1}{2}\sqrt{\frac{3}{5}}$  &$-\frac{1}{\sqrt{15}}$ &$\frac{1}{2\sqrt{15}}$ &$\frac{1}{2\sqrt{15}}$\\
$U_{32\overline{I}}$ &$\frac{1}{2}\sqrt{\frac{3}{5}}$ &$\frac{1}{2}\sqrt{\frac{3}{5}}$ &$\sqrt{\frac{3}{5}}$  &$-\frac{1}{\sqrt{15}}$ &$-\frac{1}{\sqrt{15}}$ &$\frac{1}{2\sqrt{15}}$\\
$U_{32\overline{A}}$ &$\frac{1}{2}\sqrt{\frac{3}{5}}$ &$\sqrt{\frac{3}{5}}$ &$\frac{1}{2}\sqrt{\frac{3}{5}}$  &$-\frac{1}{\sqrt{15}}$ &$\frac{1}{2\sqrt{15}}$ &$-\frac{1}{\sqrt{15}}$\\
\hline
$U_{51R}$ &$\frac{2}{\sqrt{15}}$ &$\frac{1}{2}\sqrt{\frac{5}{3}}$ &$\frac{1}{2}\sqrt{\frac{5}{3}}$  &$-\frac{1}{\sqrt{15}}$  &$-\frac{1}{2\sqrt{15}}$  &$-\frac{1}{2\sqrt{15}}$\\
$U_{51I}$ &$\frac{1}{2\sqrt{15}}$ &$\frac{1}{2}\sqrt{\frac{5}{3}}$ &$\frac{1}{\sqrt{15}}$  &$-\frac{1}{\sqrt{15}}$  &$\frac{1}{\sqrt{15}}$  &$-\frac{1}{2\sqrt{15}}$\\
$U_{51A}$ &$\frac{1}{2\sqrt{15}}$ &$\frac{1}{\sqrt{15}}$ &$\frac{1}{2}\sqrt{\frac{5}{3}}$  &$-\frac{1}{\sqrt{15}}$  &$-\frac{1}{2\sqrt{15}}$  &$\frac{1}{\sqrt{15}}$\\
\hline
$U_{51\overline{R}}$ &$\frac{1}{2}\sqrt{\frac{5}{3}}$ &$\frac{2}{\sqrt{15}}$ &$\frac{2}{\sqrt{15}}$ &$-\frac{1}{2\sqrt{15}}$ &$-\frac{1}{\sqrt{15}}$ &$-\frac{1}{\sqrt{15}}$\\
$U_{51\overline{I}}$ &$-\frac{1}{2\sqrt{15}}$ &$\frac{2}{\sqrt{15}}$ &$-\frac{1}{\sqrt{15}}$ &$-\frac{1}{2\sqrt{15}}$ &$\frac{2}{\sqrt{15}}$ &$-\frac{1}{\sqrt{15}}$\\
$U_{51\overline{A}}$ &$\frac{1}{2\sqrt{15}}$ &$\frac{1}{\sqrt{15}}$ &$-\frac{2}{\sqrt{15}}$ &$\frac{1}{2\sqrt{15}}$ &$\frac{1}{\sqrt{15}}$ &$-\frac{2}{\sqrt{15}}$\\
\hline
$U_{41IA}$ &$\sqrt{\frac{2}{5}}$ &$\sqrt{\frac{2}{5}}$ &$\frac{3}{2\sqrt{10}}$ &$-\frac{1}{2\sqrt{10}}$ &$-\frac{1}{2\sqrt{10}}$ &$-\frac{1}{\sqrt{10}}$\\
$U_{41AR}$ &$-\frac{1}{2\sqrt{10}}$ &$\frac{3}{2\sqrt{10}}$ &$-\frac{1}{2\sqrt{10}}$ &$-\frac{1}{2\sqrt{10}}$ &$\frac{3}{2\sqrt{10}}$ &$-\frac{1}{2\sqrt{10}}$\\
$U_{41RI}$ &$0$ &$\frac{1}{2\sqrt{10}}$ &$-\sqrt{\frac{2}{5}}$ &$\frac{1}{2\sqrt{10}}$ &$\frac{1}{\sqrt{10}}$ &$-\frac{3}{2\sqrt{10}}$\\
$U_{41AI}$ &$\sqrt{\frac{2}{5}}$ &$\frac{3}{2\sqrt{10}}$ &$\sqrt{\frac{2}{5}}$ &$-\frac{1}{2\sqrt{10}}$ &$-\frac{1}{\sqrt{10}}$ &$-\frac{1}{2\sqrt{10}}$\\
$U_{41RA}$ &$0$ &$\sqrt{\frac{2}{5}}$ &$-\frac{1}{2\sqrt{10}}$ &$-\frac{1}{2\sqrt{10}}$ &$\frac{3}{2\sqrt{10}}$ &$-\frac{1}{\sqrt{10}}$\\
$U_{41IR}$ &$\frac{1}{2\sqrt{10}}$ &$\frac{1}{2\sqrt{10}}$ &$-\frac{3}{2\sqrt{10}}$ &$\frac{1}{2\sqrt{10}}$ &$\frac{1}{2\sqrt{10}}$ &$-\frac{3}{2\sqrt{10}}$\\
\hline
$U_{\chi RI}$ &$\frac{3}{2\sqrt{10}}$  &$\frac{1}{2}\sqrt{\frac{5}{2}}$  &$\frac{1}{2\sqrt{10}}$  &$-\frac{1}{2\sqrt{10}}$  &$\frac{1}{2\sqrt{10}}$  &$-\frac{3}{2\sqrt{10}}$\\
$U_{\chi AR}$ &$-\frac{1}{\sqrt{10}}$ &$0$ &$-\frac{1}{2}\sqrt{\frac{5}{2}}$  &$\frac{1}{2\sqrt{10}}$  &$\frac{3}{2\sqrt{10}}$  &$-\frac{1}{\sqrt{10}}$\\
$U_{\chi IA}$ &$\frac{1}{\sqrt{10}}$ &$-\frac{1}{2\sqrt{10}}$  &$0$  &$\frac{1}{2\sqrt{10}}$  &$-\frac{1}{\sqrt{10}}$  &$-\frac{1}{2\sqrt{10}}$\\
$U_{\chi IR}$ &$\frac{1}{\sqrt{10}}$ &$\frac{1}{2}\sqrt{\frac{5}{2}}$  &$0$  &$-\frac{1}{2\sqrt{10}}$  &$\frac{1}{\sqrt{10}}$  &$-\frac{3}{2\sqrt{10}}$\\
$U_{\chi RA}$ &$\frac{3}{2\sqrt{10}}$  &$\frac{1}{2\sqrt{10}}$  &$\frac{1}{2}\sqrt{\frac{5}{2}}$  &$-\frac{1}{2\sqrt{10}}$  &$-\frac{3}{2\sqrt{10}}$  &$\frac{1}{2\sqrt{10}}$\\
$U_{\chi AI}$ &$\frac{1}{\sqrt{10}}$ &$0$ &$-\frac{1}{2\sqrt{10}}$  &$\frac{1}{2\sqrt{10}}$  &$-\frac{1}{2\sqrt{10}}$  &$-\frac{1}{\sqrt{10}}$\\
\hline
\hline
\end{tabular}
}
\egroup
\label{tab:chiral}
}
\end{table}

\begin{table}
\tbl{Models arising from the $E_6 \longrightarrow SO(10) \otimes U(1)_{42X} \longrightarrow SU(3)_C \otimes SU(2)_L \otimes G \longrightarrow G_{SM}$ chains of subgroups, where $G_{SM} \equiv SU(3)_C \otimes SU(2)_L \otimes U(1)_Y$.}
{
\bgroup                    
\def\arraystretch{2.0}
\scalebox{0.9}{
\begin{tabular}{|c|c|ccc|c|}
\hline
\hline  
Model      &$G$ factor                                            &                      &                      &                     & \\          
\hline
$A1_{XY}$  &$U(1)_{32Y}\otimes U(1)_{\chi XY}\otimes U(1)_{42X}$  &$k_a$                 &$k_b$                 &$k_c$                & \\
\hline
$A1_{IR}$: &$U(1)_{32R}\otimes U(1)_{\chi IR}\otimes U(1)_{42I}$  &$-\frac{1}{\sqrt{15}}$&$\frac{1}{\sqrt{10}}$ &$-\sqrt{\frac{3}{2}}$& \\                
$A1_{AR}$: &$U(1)_{32R}\otimes U(1)_{\chi AR}\otimes U(1)_{42A}$  &$-\frac{1}{\sqrt{15}}$&$2\sqrt{\frac{2}{5}}$ &$0$                  & \\
$A1_{RI}$: &$U(1)_{32I}\otimes U(1)_{\chi RI}\otimes U(1)_{42R}$  &$\sqrt{\frac{5}{3}}$  &$0$                   &$0$                  &~\cite{Georgi:1974sy,Barr:1981qv,London:1986dk}\\
$A1_{AI}$: &$U(1)_{32I}\otimes U(1)_{\chi AI}\otimes U(1)_{42A}$  &$\sqrt{\frac{5}{3}}$  &$0$                   &$0$                  &~\cite{King:2005jy} \\
$A1_{RA}$: &$U(1)_{32A}\otimes U(1)_{\chi RA}\otimes U(1)_{42R}$  &$\frac{1}{\sqrt{15}}$ &$-2\sqrt{\frac{2}{5}}$&$0$                  &~\cite{Barr:1981qv} \\
$A1_{IA}$: &$U(1)_{32A}\otimes U(1)_{\chi IA}\otimes U(1)_{42I}$  &$\frac{1}{\sqrt{15}}$ &$-\frac{1}{\sqrt{10}}$&$-\sqrt{\frac{3}{2}}$& \\
\hline
$A2_{X}$   &$U(1)_{X}\otimes U(1)_{31X}\otimes U(1)_{42X}$        &$k_a$                 &$k_b$                 &$k_c$                &\\
\hline
$A2_{R}$:  &$U(1)_{R}\otimes U(1)_{31R}\otimes U(1)_{42R}$        &$1$                   &$-\sqrt{\frac{2}{3}}$ &$0$                  &~\cite{Pati:1974yy,Mohapatra:1974hk}\\                
$A2_{I}$:  &$U(1)_{I}\otimes U(1)_{31I}\otimes U(1)_{42I}$        &$0$                   &$\frac{1}{\sqrt{6}}$  &$-\sqrt{\frac{3}{2}}$&~\cite{Rodriguez:2016cgr}\\
$A2_{A}$:  &$U(1)_{A}\otimes U(1)_{31A}\otimes U(1)_{42A}$        &$-1$                  &$\sqrt{\frac{2}{3}}$  &$0$                  &~\cite{Ma:1986we}\\
\hline
$A3_{X}$   &$U(1)_{31X} \otimes U(1)_{42X}$                       &$k_a$                 &$k_b$                 &                     & \\
\hline
$A3_{R}$:  &$U(1)_{31R}\otimes U(1)_{42R}$                        &$--$                  &$--$                  &                     & \\                
$A3_{I}$:  &$U(1)_{31I}\otimes U(1)_{42I}$                        &$\frac{1}{\sqrt{6}}$  &$-\sqrt{3/2}$         &                     & \\
$A3_{A}$:  &$U(1)_{31A}\otimes U(1)_{42A}$                        &$--$                  &$--$                  &                     & \\
\hline
\hline
\end{tabular}
}
\egroup
\label{table:a}
}
\end{table}

\begin{table}
\tbl{Models arising from the $E_6 \longrightarrow SU(2)_X \otimes SU(6) \longrightarrow SU(3)_C \otimes SU(2)_L \otimes G \longrightarrow G_{SM}$ chains of subgroups, where $G_{SM} \equiv SU(3)_C \otimes SU(2)_L \otimes U(1)_Y$.}
{
\bgroup                    
\def\arraystretch{2.0}
\begin{tabular}{|c|c|ccc|c|}
\hline
\hline  
Model    &$G$ factor                                             &                      &                     &     &\\          
\hline
$B1_{X}$ &$U(1)_{32X}\otimes U(1)_{51X} \otimes U(1)_{X}$        &$k_a$                 &$k_b$                &$k_c$&\\
\hline
$B1_{R}$:&$U(1)_{32R}\otimes U(1)_{51R}\otimes U(1)_{R}$         &$-\frac{1}{\sqrt{15}}$&$-\sqrt{\frac{3}{5}}$&$1$  &\\                
$B1_{I}$:&$U(1)_{32I}\otimes U(1)_{51I}\otimes U(1)_{I}$         &$\sqrt{\frac{5}{3}}$  &$0$                  &$0$  &~\cite{Witten:1985xc}\\
$B1_{A}$:&$U(1)_{32A}\otimes U(1)_{51A}\otimes U(1)_{A}$         &$\frac{1}{\sqrt{15}}$ &$-\sqrt{\frac{3}{5}}$&$-1$ &\\
\hline
$B2_{X}$ &$U(1)_{31X}\otimes U(1)_{42X}\otimes U(1)_{X}$         &\ \ \  same as        &$A2_{X}$             &     &\\
\hline
$B3_{X}$ &$U(1)_{33}\otimes U(1)_{21\overline{X}}\otimes U(1)_X$ &\ \ \  same as        &$D_X$                &    &  \\
\hline
\hline
\end{tabular}
\egroup
\label{table:b}
}
\end{table}

\begin{table}
\tbl{Models arising from the $E_6 \longrightarrow SU(2)_L \otimes SU(6) \longrightarrow SU(3)_C \otimes SU(2)_L \otimes G \longrightarrow G_{SM}$ chains of subgroups, where $G_{SM} \equiv SU(3)_C \otimes SU(2)_L \otimes U(1)_Y$.}
{
\bgroup                    
\def\arraystretch{2.0}
\begin{tabular}{|c|c|ccc|c|}
\hline
\hline  
Model     &$G$ factor                                                          &                     &                      &        \\          
\hline
$C1_{XY}$ &$U(1)_{31Y}\otimes U(1)_{41XY}\otimes U(1)_{51\overline{X}}$        &$k_a$                &$k_b$                 &$k_c$ \\
\hline
$C1_{IR}$:&$U(1)_{31R}\otimes U(1)_{41IR}\otimes U(1)_{51\overline{I}}$        &$-\sqrt{\frac{2}{3}}$&$\sqrt{\frac{2}{5}}$  &$\sqrt{\frac{3}{5}}$ \\                
$C1_{AR}$:&$U(1)_{31R}\otimes U(1)_{41AR}\otimes U(1)_{51\overline{A}}$        &$-\sqrt{\frac{2}{3}}$&$\sqrt{\frac{2}{5}}$  &$\sqrt{\frac{3}{5}}$ \\
$C1_{RI}$:&$U(1)_{31I}\otimes U(1)_{41RI}\otimes U(1)_{51\overline{R}}$        &$\frac{1}{\sqrt{6}}$ &$\frac{3}{\sqrt{10}}$ &$-\sqrt{\frac{3}{5}}$\\
$C1_{AI}$:&$U(1)_{31I}\otimes U(1)_{41AI}\otimes U(1)_{51\overline{A}}$        &$\frac{1}{\sqrt{6}}$ &$-\frac{3}{\sqrt{10}}$&$\sqrt{\frac{3}{5}}$\\                
$C1_{RA}$:&$U(1)_{31A}\otimes U(1)_{41RA}\otimes U(1)_{51\overline{R}}$        &$\sqrt{\frac{2}{3}}$ &$\sqrt{\frac{2}{5}}$  &$-\sqrt{\frac{3}{5}}$ \\
$C1_{IA}$:&$U(1)_{31A}\otimes U(1)_{41IA}\otimes U(1)_{51\overline{I}}$        &$\sqrt{\frac{2}{3}}$ &$-\sqrt{\frac{2}{5}}$ &$\sqrt{\frac{3}{5}}$ \\
\hline
$C2_{X}$  &$U(1)_{X}\otimes U(1)_{32\overline{X}}\otimes U(1)_{51\overline{X}}$&$k_a$                &$k_b$                 &$k_c$ \\
\hline
$C2_{R}$: &$U(1)_{R}\otimes U(1)_{32\overline{R}}\otimes U(1)_{51\overline{R}}$&$1$                  &$-\frac{1}{\sqrt{15}}$&$-\sqrt{\frac{3}{5}}$ \\                
$C2_{I}$: &$U(1)_{I}\otimes U(1)_{32\overline{I}}\otimes U(1)_{51\overline{I}}$&$0$                  &$-\frac{4}{\sqrt{15}}$&$\sqrt{\frac{3}{5}}$ \\
$C2_{A}$: &$U(1)_{A}\otimes U(1)_{32\overline{A}}\otimes U(1)_{51\overline{A}}$&$-1$                 &$-\frac{1}{\sqrt{15}}$&$\sqrt{\frac{3}{5}}$  \\
\hline
$C3_{X}$  &$U(1)_{X}\otimes U(1)_{31X}\otimes U(1)_{42X}$                      &\ \ \  same as       &$A2_{X}$              &     \\  
\hline
$C4_{X}$   &$U(1)_X \otimes U(1)_{21\overline{X}}\otimes  U(1)_{33}$            &\ \ \  same as       &$D_X$                 &      \\
\hline
\hline
\end{tabular}
\egroup
\label{table:c}
}
\end{table}

\begin{table}
\tbl{Models arising from the $E_6 \longrightarrow SU(3) \otimes SU(3) \otimes SU(3) \longrightarrow SU(3)_C \otimes SU(2)_L \otimes G \longrightarrow G_{SM}$ chains of subgroups, where $G_{SM} \equiv SU(3)_C \otimes SU(2)_L \otimes U(1)_Y$.}
{
\bgroup                    
\def\arraystretch{2.0}
\begin{tabular}{|c|c|ccc|c|}
\hline
\hline  
Model   &$G$ factor                                              &     &                     &                     &\\          
\hline
$D_{X}$ &$U(1)_X \otimes U(1)_{21\overline{X}}\otimes U(1)_{33}$ &$k_a$&$k_b$                &$k_c$                &\\
\hline
$D_{R}$:&$U(1)_R \otimes U(1)_{21\overline{R}}\otimes  U(1)_{33}$&$1$  &$-\frac{1}{\sqrt{3}}$&$-\frac{1}{\sqrt{3}}$& \\                
$D_{I}$:&$U(1)_I \otimes U(1)_{21\overline{I}}\otimes  U(1)_{33}$&$0$  &$\frac{2}{\sqrt{3}}$ &$-\frac{1}{\sqrt{3}}$&~\cite{Achiman:1977py,Achiman:1978vg,Glashow:1984gc,Babu:1985gi,Rodriguez:2016cgr}\\
$D_{A}$:&$U(1)_A \otimes U(1)_{21\overline{A}}\otimes  U(1)_{33}$&$-1$ &$\frac{1}{\sqrt{3}}$ &$-\frac{1}{\sqrt{3}}$& \\
\hline
\hline
\end{tabular}
\egroup
\label{table:d}
}
\end{table}


 \FloatBarrier
%

\bibliographystyle{ws-ijmpa}
\bibliography{referencese6}

\begin{thebibliography}{10}
\expandafter\ifx\csname urlstyle\endcsname\relax
  \providecommand{\doi}[1]{doi:\discretionary{}{}{}#1}\else
  \providecommand{\doi}{doi:\discretionary{}{}{}\begingroup
  \urlstyle{rm}\Url}\fi

\bibitem{Barr:1981qv}
S.~M. Barr, {\em Phys. Lett.} {\bf 112B}, 219  (1982),
  \doi{10.1016/0370-2693(82)90966-2}.

\bibitem{Robinett:1982tq}
R.~W. Robinett and J.~L. Rosner, {\em Phys. Rev.} {\bf D26},   2396  (1982),
  \doi{10.1103/PhysRevD.26.2396}.

\bibitem{Witten:1985xc}
E.~Witten, {\em Nucl. Phys.} {\bf B258},  ~75  (1985),
  \doi{10.1016/0550-3213(85)90603-0}.

\bibitem{Ma:1986we}
E.~Ma, {\em Phys. Rev.} {\bf D36},   274  (1987),
  \doi{10.1103/PhysRevD.36.274}.

\bibitem{Ma:1995xk}
E.~Ma, {\em Phys. Lett.} {\bf B380}, 286  (1996),
  \href{http://arxiv.org/abs/hep-ph/9507348}{{\ttfamily arXiv:hep-ph/9507348
  [hep-ph]}}, \doi{10.1016/0370-2693(96)00524-2}.

\bibitem{Martinez:2001mu}
R.~Martinez, W.~A. Ponce and L.~A. Sanchez, {\em Phys. Rev.} {\bf D65},
  055013  (2002), \href{http://arxiv.org/abs/hep-ph/0110246}{{\ttfamily
  arXiv:hep-ph/0110246 [hep-ph]}}, \doi{10.1103/PhysRevD.65.055013}.

\bibitem{Rojas:2015tqa}
E.~Rojas and J.~Erler, {\em JHEP} {\bf 10},   063  (2015),
  \href{http://arxiv.org/abs/1505.03208}{{\ttfamily arXiv:1505.03208
  [hep-ph]}}, \doi{10.1007/JHEP10(2015)063}.

\bibitem{Mantilla:2016sew}
S.~F. Mantilla, R.~Martinez, F.~Ochoa and C.~F. Sierra, {\em Nucl. Phys.} {\bf
  B911}, 338  (2016), \href{http://arxiv.org/abs/1602.05216}{{\ttfamily
  arXiv:1602.05216 [hep-ph]}}, \doi{10.1016/j.nuclphysb.2016.08.014}.

\bibitem{Huang:2017uli}
C.-S. Huang, W.-J. Li and X.-H. Wu  (2017),
  \href{http://arxiv.org/abs/1705.01411}{{\ttfamily arXiv:1705.01411
  [hep-ph]}}.

\bibitem{CarcamoHernandez:2017owh}
A.~E. Cárcamo~Hernández, S.~Kovalenko, J.~W.~F. Valle and C.~A.
  Vaquera-Araujo, {\em JHEP} {\bf 07},   118  (2017),
  \href{http://arxiv.org/abs/1705.06320}{{\ttfamily arXiv:1705.06320
  [hep-ph]}}, \doi{10.1007/JHEP07(2017)118}.

\bibitem{Derendinger:1983aj}
J.~P. Derendinger, J.~E. Kim and D.~V. Nanopoulos, {\em Phys. Lett.} {\bf
  139B}, 170  (1984), \doi{10.1016/0370-2693(84)91238-3}.

\bibitem{King:2005jy}
S.~F. King, S.~Moretti and R.~Nevzorov, {\em Phys. Rev.} {\bf D73},   035009
  (2006), \href{http://arxiv.org/abs/hep-ph/0510419}{{\ttfamily
  arXiv:hep-ph/0510419 [hep-ph]}}, \doi{10.1103/PhysRevD.73.035009}.

\bibitem{Rodriguez:2016cgr}
O.~Rodríguez, R.~H. Benavides, W.~A. Ponce and E.~Rojas, {\em Phys. Rev.} {\bf
  D95},   014009  (2017), \href{http://arxiv.org/abs/1605.00575}{{\ttfamily
  arXiv:1605.00575 [hep-ph]}}, \doi{10.1103/PhysRevD.95.014009}.

\bibitem{Dimopoulos:1985xs}
S.~Dimopoulos and L.~J. Hall, {\em Nucl. Phys.} {\bf B255}, 633  (1985),
  \doi{10.1016/0550-3213(85)90157-9}.

\bibitem{Rizos:1997pe}
J.~Rizos and K.~Tamvakis, {\em Phys. Lett.} {\bf B414}, 277  (1997),
  \href{http://arxiv.org/abs/hep-ph/9702295}{{\ttfamily arXiv:hep-ph/9702295
  [hep-ph]}}, \doi{10.1016/S0370-2693(97)01181-7}.

\bibitem{Shafi:1998jf}
Q.~Shafi and Z.~Tavartkiladze, {\em Nucl. Phys.} {\bf B552}, 67  (1999),
  \href{http://arxiv.org/abs/hep-ph/9807502}{{\ttfamily arXiv:hep-ph/9807502
  [hep-ph]}}, \doi{10.1016/S0550-3213(99)00178-9}.

\bibitem{Faraggi:2014bla}
A.~E. Faraggi, M.~Paraskevas, J.~Rizos and K.~Tamvakis, {\em Phys. Rev.} {\bf
  D90},   015036  (2014), \href{http://arxiv.org/abs/1405.2274}{{\ttfamily
  arXiv:1405.2274 [hep-ph]}}, \doi{10.1103/PhysRevD.90.015036}.

\bibitem{Dong:2017zxo}
P.~V. Dong, D.~T. Huong, F.~S. Queiroz, J.~W.~F. Valle and C.~A. Vaquera-Araujo
   (2017), \href{http://arxiv.org/abs/1710.06951}{{\ttfamily arXiv:1710.06951
  [hep-ph]}}.

\bibitem{Benavides:2016utf}
R.~Benavides, L.~A. Muñoz, W.~A. Ponce, O.~Rodríguez and E.~Rojas, {\em Phys.
  Rev.} {\bf D95},   115018  (2017),
  \href{http://arxiv.org/abs/1612.07660}{{\ttfamily arXiv:1612.07660
  [hep-ph]}}, \doi{10.1103/PhysRevD.95.115018}.

\bibitem{Langacker:2008yv}
P.~Langacker, {\em Rev. Mod. Phys.} {\bf 81}, 1199  (2009),
  \href{http://arxiv.org/abs/0801.1345}{{\ttfamily arXiv:0801.1345 [hep-ph]}},
  \doi{10.1103/RevModPhys.81.1199}.

\bibitem{Salazar:2015gxa}
C.~Salazar, R.~H. Benavides, W.~A. Ponce and E.~Rojas, {\em JHEP} {\bf 07},
  096  (2015), \href{http://arxiv.org/abs/1503.03519}{{\ttfamily
  arXiv:1503.03519 [hep-ph]}}, \doi{10.1007/JHEP07(2015)096}.

\bibitem{Gursey:1975ki}
F.~Gursey, P.~Ramond and P.~Sikivie, {\em Phys. Lett.} {\bf 60B}, 177  (1976),
  \doi{10.1016/0370-2693(76)90417-2}.

\bibitem{Achiman:1978vg}
Y.~Achiman and B.~Stech, {\em Phys. Lett.} {\bf 77B}, 389  (1978),
  \doi{10.1016/0370-2693(78)90584-1}.

\bibitem{London:1986dk}
D.~London and J.~L. Rosner, {\em Phys. Rev.} {\bf D34},   1530  (1986),
  \doi{10.1103/PhysRevD.34.1530}.

\bibitem{Camargo-Molina:2017kxd}
J.~E. Camargo-Molina, A.~P. Morais, A.~Ordell, R.~Pasechnik and J.~Wessén
  (2017), \href{http://arxiv.org/abs/1711.05199}{{\ttfamily arXiv:1711.05199
  [hep-ph]}}.

\bibitem{Erler:2000wu}
J.~Erler, {\em Nucl. Phys.} {\bf B586}, 73  (2000),
  \href{http://arxiv.org/abs/hep-ph/0006051}{{\ttfamily arXiv:hep-ph/0006051
  [hep-ph]}}, \doi{10.1016/S0550-3213(00)00427-2}.

\bibitem{Carena:2004xs}
M.~Carena, A.~Daleo, B.~A. Dobrescu and T.~M.~P. Tait, {\em Phys. Rev.} {\bf
  D70},   093009  (2004), \href{http://arxiv.org/abs/hep-ph/0408098}{{\ttfamily
  arXiv:hep-ph/0408098 [hep-ph]}}, \doi{10.1103/PhysRevD.70.093009}.

\bibitem{Erler:2011ud}
J.~Erler, P.~Langacker, S.~Munir and E.~Rojas, {\em JHEP} {\bf 11},   076
  (2011), \href{http://arxiv.org/abs/1103.2659}{{\ttfamily arXiv:1103.2659
  [hep-ph]}}, \doi{10.1007/JHEP11(2011)076}.

\bibitem{Slansky:1981yr}
R.~Slansky, {\em Phys. Rept.} {\bf 79}, 1  (1981),
  \doi{10.1016/0370-1573(81)90092-2}.

\bibitem{Erler:2009jh}
J.~Erler, P.~Langacker, S.~Munir and E.~Rojas, {\em JHEP} {\bf 08},   017
  (2009), \href{http://arxiv.org/abs/0906.2435}{{\ttfamily arXiv:0906.2435
  [hep-ph]}}, \doi{10.1088/1126-6708/2009/08/017}.

\bibitem{Holdom:1985ag}
B.~Holdom, {\em Phys. Lett.} {\bf 166B}, 196  (1986),
  \doi{10.1016/0370-2693(86)91377-8}.

\bibitem{Babu:1996vt}
K.~S. Babu, C.~F. Kolda and J.~March-Russell, {\em Phys. Rev.} {\bf D54}, 4635
  (1996), \href{http://arxiv.org/abs/hep-ph/9603212}{{\ttfamily
  arXiv:hep-ph/9603212 [hep-ph]}}, \doi{10.1103/PhysRevD.54.4635}.

\bibitem{Babu:1997st}
K.~S. Babu, C.~F. Kolda and J.~March-Russell, {\em Phys. Rev.} {\bf D57}, 6788
  (1998), \href{http://arxiv.org/abs/hep-ph/9710441}{{\ttfamily
  arXiv:hep-ph/9710441 [hep-ph]}}, \doi{10.1103/PhysRevD.57.6788}.

\bibitem{delAguila:1988jz}
F.~del Aguila, G.~D. Coughlan and M.~Quiros, {\em Nucl. Phys.} {\bf B307},
  633  (1988), \doi{10.1016/0550-3213(88)90266-0}, [Erratum: Nucl.
  Phys.B312,751(1989)].

\bibitem{Baulieu:1981ux}
L.~Baulieu and R.~Coquereaux, {\em Annals Phys.} {\bf 140},   163  (1982),
  \doi{10.1016/0003-4916(82)90339-6}.

\bibitem{Georgi:1974sy}
H.~Georgi and S.~L. Glashow, {\em Phys. Rev. Lett.} {\bf 32}, 438  (1974),
  \doi{10.1103/PhysRevLett.32.438}.

\bibitem{Erler:2009ut}
J.~Erler, P.~Langacker, S.~Munir and E.~Rojas, {\em AIP Conf. Proc.} {\bf
  1200}, 790  (2010), \href{http://arxiv.org/abs/0910.0269}{{\ttfamily
  arXiv:0910.0269 [hep-ph]}}, \doi{10.1063/1.3327731}.

\bibitem{Erler:2010uy}
J.~Erler, P.~Langacker, S.~Munir and E.~rojas, { Z' searches: From tevatron to
  lhc}, in {\em 22nd Rencontres de Blois on Particle Physics and Cosmology
  Blois, Loire Valley, France, July 15-20, 2010\/}, .

\bibitem{Erler:2011iw}
J.~Erler, P.~Langacker, S.~Munir and E.~Rojas, { {Z' Bosons from E(6): Collider
  and Electroweak Constraints}}, in {\em {19th International Workshop on
  Deep-Inelastic Scattering and Related Subjects (DIS 2011) Newport News,
  Virginia, April 11-15, 2011}\/},  (2011).
\newblock \href{http://arxiv.org/abs/1108.0685}{{\ttfamily arXiv:1108.0685
  [hep-ph]}}.

\bibitem{Sanchez:2001ua}
L.~A. Sanchez, W.~A. Ponce and R.~Martinez, {\em Phys. Rev.} {\bf D64},
  075013  (2001), \href{http://arxiv.org/abs/hep-ph/0103244}{{\ttfamily
  arXiv:hep-ph/0103244 [hep-ph]}}, \doi{10.1103/PhysRevD.64.075013}.

\bibitem{Pati:1974yy}
J.~C. Pati and A.~Salam, {\em Phys. Rev.} {\bf D10}, 275  (1974),
  \doi{10.1103/PhysRevD.10.275, 10.1103/PhysRevD.11.703.2}, [Erratum: Phys.
  Rev.D11,703(1975)].

\bibitem{Glashow:1961tr}
S.~L. Glashow, {\em Nucl. Phys.} {\bf 22}, 579  (1961),
  \doi{10.1016/0029-5582(61)90469-2}.

\bibitem{Weinberg:1967tq}
S.~Weinberg, {\em Phys. Rev. Lett.} {\bf 19}, 1264  (1967),
  \doi{10.1103/PhysRevLett.19.1264}.

\bibitem{Erler:2002pr}
J.~Erler, P.~Langacker and T.-j. Li, {\em Phys. Rev.} {\bf D66},   015002
  (2002), \href{http://arxiv.org/abs/hep-ph/0205001}{{\ttfamily
  arXiv:hep-ph/0205001 [hep-ph]}}, \doi{10.1103/PhysRevD.66.015002}.

\bibitem{Kang:2004pp}
J.~Kang, P.~Langacker, T.-j. Li and T.~Liu, {\em Phys. Rev. Lett.} {\bf 94},
  061801  (2005), \href{http://arxiv.org/abs/hep-ph/0402086}{{\ttfamily
  arXiv:hep-ph/0402086 [hep-ph]}}, \doi{10.1103/PhysRevLett.94.061801}.

\bibitem{Singer:1980sw}
M.~Singer, J.~W.~F. Valle and J.~Schechter, {\em Phys. Rev.} {\bf D22},   738
  (1980), \doi{10.1103/PhysRevD.22.738}.

\bibitem{Mohapatra:1974hk}
R.~N. Mohapatra and J.~C. Pati, {\em Phys. Rev.} {\bf D11}, 566  (1975),
  \doi{10.1103/PhysRevD.11.566}.

\bibitem{Achiman:1977py}
Y.~Achiman, {\em Phys. Lett.} {\bf 70B}, 187  (1977),
  \doi{10.1016/0370-2693(77)90517-2}.

\bibitem{Glashow:1984gc}
S.~L. Glashow, { {Trinification of All Elementary Particle Forces}}, in {\em
  {Fifth Workshop on Grand Unification Providence, Rhode Island, April 12-14,
  1984}\/},  (1984), p. 0088.

\bibitem{Babu:1985gi}
K.~S. Babu, X.-G. He and S.~Pakvasa, {\em Phys. Rev.} {\bf D33},   763  (1986),
  \doi{10.1103/PhysRevD.33.763}.

\bibitem{Bai:2017zhj}
Y.~Bai and B.~A. Dobrescu  (2017),
  \href{http://arxiv.org/abs/1710.01456}{{\ttfamily arXiv:1710.01456
  [hep-ph]}}.

\bibitem{Aaboud:2017buh}
 ATLAS Collaboration (M.~Aaboud {\em et~al.}), {\em JHEP} {\bf 10},   182
  (2017), \href{http://arxiv.org/abs/1707.02424}{{\ttfamily arXiv:1707.02424
  [hep-ex]}}, \doi{10.1007/JHEP10(2017)182}.

\bibitem{ATLAS:2016cyf}
 ATLAS Collaboration (T.~A. collaboration)  (2016).

\bibitem{Hati:2015awg}
C.~Hati, G.~Kumar and N.~Mahajan, {\em JHEP} {\bf 01},   117  (2016),
  \href{http://arxiv.org/abs/1511.03290}{{\ttfamily arXiv:1511.03290
  [hep-ph]}}, \doi{10.1007/JHEP01(2016)117}.

\bibitem{Joglekar:2016yap}
A.~Joglekar and J.~L. Rosner, {\em Phys. Rev.} {\bf D96},   015026  (2017),
  \href{http://arxiv.org/abs/1607.06900}{{\ttfamily arXiv:1607.06900
  [hep-ph]}}, \doi{10.1103/PhysRevD.96.015026}.

\bibitem{Das:2016vkr}
D.~Das, C.~Hati, G.~Kumar and N.~Mahajan, {\em Phys. Rev.} {\bf D94},   055034
  (2016), \href{http://arxiv.org/abs/1605.06313}{{\ttfamily arXiv:1605.06313
  [hep-ph]}}, \doi{10.1103/PhysRevD.94.055034}.

\bibitem{Dorsner:2017ufx}
I.~Doršner, S.~Fajfer, D.~A. Faroughy and N.~Košnik  (2017),
  \href{http://arxiv.org/abs/1706.07779}{{\ttfamily arXiv:1706.07779
  [hep-ph]}}, \doi{10.1007/JHEP10(2017)188}, [JHEP10,188(2017)].

\bibitem{Blanke:2018sro}
M.~Blanke and A.~Crivellin  (2018),
  \href{http://arxiv.org/abs/1801.07256}{{\ttfamily arXiv:1801.07256
  [hep-ph]}}.

\bibitem{Popov:2016fzr}
O.~Popov and G.~A. White, {\em Nucl. Phys.} {\bf B923}, 324  (2017),
  \href{http://arxiv.org/abs/1611.04566}{{\ttfamily arXiv:1611.04566
  [hep-ph]}}, \doi{10.1016/j.nuclphysb.2017.08.007}.

\end{thebibliography}

\end{document}